\documentclass[aps,prl,onecolumn,groupedaddress]{revtex4}
\usepackage{graphicx}
\usepackage{epstopdf}
\usepackage{amsmath}

\usepackage{amsmath,amsthm,amssymb,euscript,mathrsfs,epsf,epsfig}

\def\a{\alpha} \def\b{\beta}   
   \def\q{\theta}
    \def\m{\mu}
  \def\p{\pi}  
 \def\s{\sigma}   \def\f{\varphi}
   \def\w{\omega}

  \def\P{\Pi}  
   
   \def\W{\Omega}

\def\fr{\frac}

\begin{document}

\title{Unified phonon-based approach to the thermodynamics of solid, liquid and gas states}

\author{Dima Bolmatov$^{1}$}
\email{d.bolmatov@gmail.com, bolmatov@bnl.gov}
\author{Dmitry Zav'yalov$^{2}$}
\author{Mikhail Zhernenkov$^{1}$}
\author{Edvard T. Musaev$^{3}$}
\author{Yong Q. Cai$^{1}$}

\affiliation{$^1$ National Synchrotron Light Source II, Brookhaven National Laboratory, Upton, NY 11973, USA}
\affiliation{$^2$ Volgograd State Technical University, Volgograd, 400005 Russia}
\affiliation{$^3$ National Research University Higher School of Economics, Faculty of Mathematics, Moscow  117312, Russia}

\begin{abstract}
We introduce an unified approach to states of matter (solid, liquid and gas) and describe the thermodynamics of the pressure-temperature phase diagram in terms of phonon excitations. We derive the effective Hamiltonian with low-energy cutoff in two transverse phonon polarisations (phononic band gaps) by breaking the symmetry in phonon interactions. Further, we construct the statistical mechanics of states of aggregation  employing the Debye approximation. The introduced formalism covers the Debye theory of solids, the phonon theory of liquids, and thermodynamic limits such as the Delong-Petit thermodynamic limit ($c_V=3k_{\rm B}$), the ideal gas limit ($c_V=\frac{3}{2}k_{\rm B}$) and the new thermodynamic limit ($c_V=2k_{\rm B}$), dubbed here the Frenkel line thermodynamic limit. We discuss the phonon propagation and localisation effects in liquids above and below the Frenkel line, and explain the "fast sound" phenomenon. As a test for our theory we calculate velocity-velocity autocorrelation and pair distribution functions within the Green-Kubo formalism. We show the consistency between dynamics of phonons and pair correlations  in the framework of the unified approach. New directions towards advancements in phononic band gaps engineering, hypersound manipulation technologies and exploration of exotic behaviour of fluids relevant to geo- and planetary sciences are discussed. The presented results are equally important both for practical implications and for fundamental research. \\

*email: bolmatov@bnl.gov
\end{abstract}

\maketitle
\section{Introduction}

Natural sciences in general are based primarily on experiment, and, what is more, on quantitative experiment. However, no series experiments can constitute a theory until a rigorous logical relationship is established between them. Theory not only allows us to systematize the available experimental results, but also makes it possible to predict new facts which can be experimentally verified \cite{llandau}. Statistical mechanics is very prominent part of theoretical physics. It forms a separate field which includes the theory of phase transitions and is closely related with the more mathematical ergodic theory and some parts of probability theory. In the annals of statistical mechanics, the last century and recent decades mark a very vibrant epoch \cite{aeinstein,pdebye,vwaals,mborn,ffrenkel,jpercus,glasinio,jlebowitz,bwidom,mfisher,kwilson,
lkadanoff,nashcroft,parisi,ruffo,crooks,dyre1,england}.  

Statistical mechanics is the art of predicting the behavior of a system with a large number of degrees of freedom, given the laws governing its microscopic behavior. The statistical description of liquids, in comparison with the solid and gas phases, was incomplete for a long time \cite{granat}.  Due to the simultaneous presence of strong interactions and large atomic displacements, common models and approximations used for gases and solids, do not apply to liquids. For this reason, liquids do not generally fall into any simple classiﬁcation, and have as a result been mostly treated as general many-body
systems. Therefore, the problem of formulating a rigorous mathematical description of all basic states of matter such as solids, liquids and gases based on a unified approach has always been regarded as illusive.

\begin{figure} 
	\centering
\includegraphics[scale=0.08]{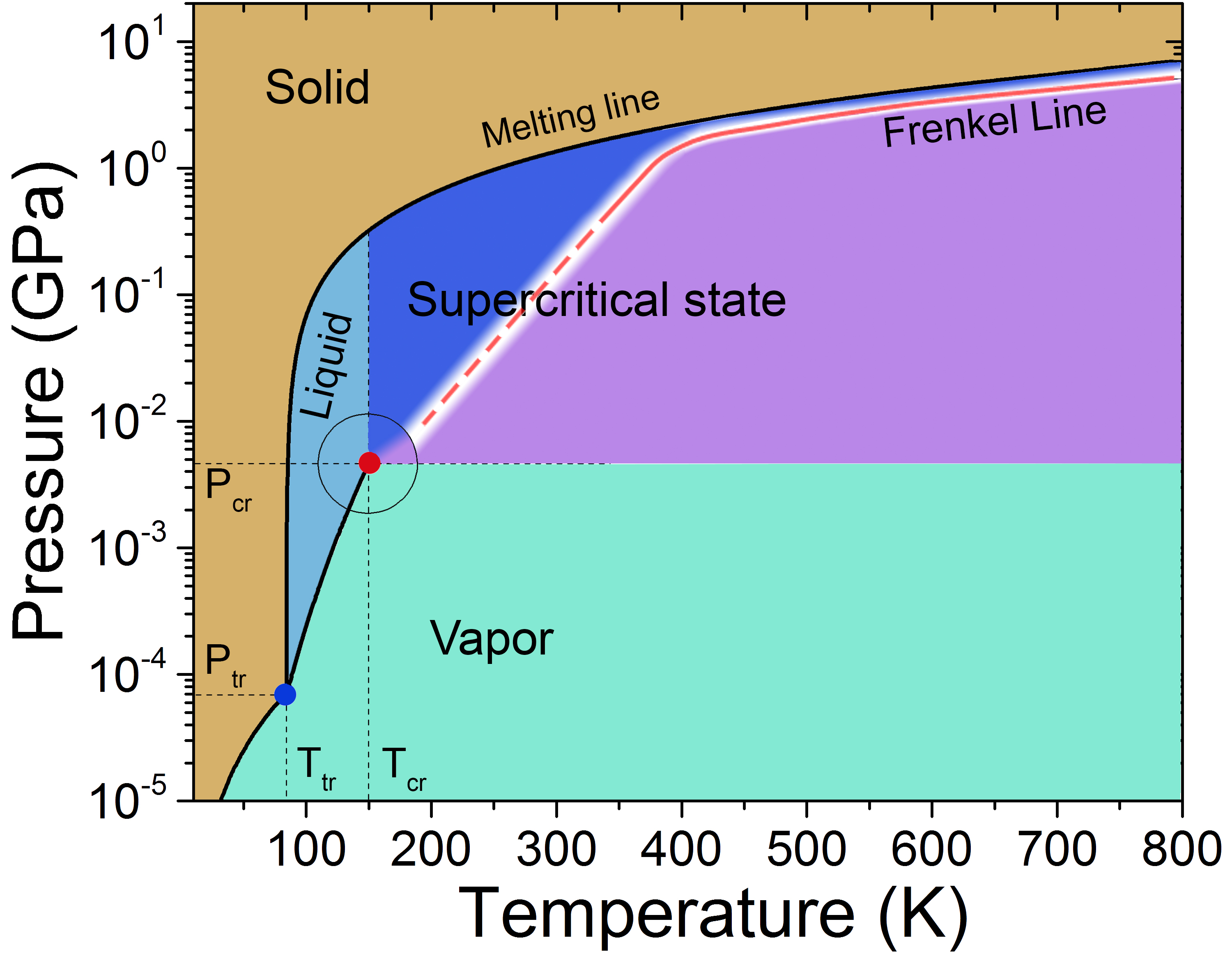}
\caption{Argon P-T phase diagram with the new thermodynamic boundary, dubbed the Frenkel line. The phase diagram shows all major states of matter.}
	\label{fig1}
\end{figure}

In this work, we introduce the {\it phonon  Hamiltonian} with an interaction part reflecting the fact that interacting phonons can be created or annihilated  making materials anharmonic over the heat production. We derive an effective Hamiltonian with two naturally emerged transverse phononic band gaps as a result of symmetry breaking in phonon interactions. The emergence of two transverse phononic gaps in a liquid spectrum rigorously proves  the Frenkel's phenomenological microscopic picture about phonon excitations  \cite{ffrenkel}. Further, we construct the statistical mechanics of states of matter existing on the pressure-temperature (P-T) phase diagram (see Fig. (\ref{fig1})) by employing the Debye approximation. The introduced formalism covers the continuous theory of solids by Debye \cite{pdebye}, the phonon theory of liquids and supercritical fluids, the well-known thermodynamic limits such as the Delong-Petit law ($c_V=3k_{\rm B}$) and the ideal gas limit ($c_V=\frac{3}{2}k_{\rm B}$), and we derive the new thermodynamic limit ($c_V=2k_{\rm B}$), dubbed here the Frenkel line thermodynamic limit. In the framework of the phonon theory of liquids we discuss the phonon propagation and localisation effects in disordered phases above and below the Frenkel line, and explain the "fast sound" phenomenon. As a test for our theory we calculate velocity-velocity autocorrelation (VACF) and pair distribution functions $g(r)$ in the framework of the linear response Green-Kubo phenomenological formalism. We  show the consistency between dynamics (phonon excitations and phononic gaps), structure (pair correlations) and thermodynamics (heat capacity $c_V$) on the P-T phase diagram (see Fig. (\ref{fig1})) within the introduced formalism. This point is of prime significance, since the manipulation of phononic band gaps in the so-called phononic crystals \cite{maldovan} is customarily realized through structural engineering. This is of crucial importance to implement heat management, since heat transfer in insulators uses phonons as carriers. In this work, it is shown
that the phononic gap manipulation can be achieved even
in simplest materials through a suitable tuning of thermodynamic
conditions. This reflects into a tuning of the atomic dynamics. Tuning of the atomic dynamics and thermally triggered localization of low- and high-frequency propagating phonon modes are the key factors for the advancement of technologies based on the sound control and manipulation at the terahertz regime. 
Finally, we outline new directions towards future advancements in the scope of sound control technologies and THz phononic band gaps engineering.

\section{Collective excitations}
\subsection{Emergence of transverse phononic gaps}

For most cases of interest, dynamics of a complex condensed matter system can be effectively described by collective excitations of quantised lattice vibrations, the so-called phonons. In the most general form the corresponding Hamiltonian can be written as $H=H_{0}+H_{int}$, where $H_{0}$ defines a free theory
\begin{equation}
\label{H_0}
H_{0}=\frac{1}{2}\sum_{0\leq\hat{\w}_q\leq\omega_{\rm D}}\left[\Pi_q^\a\Pi_{-q}^\a+\mu\hat{\w}_q^2 Q_q^\a Q_{-q}^\a\right]
\end{equation}
and the interaction potential $H_{int}$ is of higher order in fields. The coefficient $\hat{\w}_q^2$ here gives the dispersion relation of a phonon and the additional dimensionless prefactor $\m$ will be needed further to identify the states of matter (see Table I). The hat emphasises that the resulting frequency for longitudinal and transverse modes will be different after symmetry breaking.

 Following the construction \cite{sci1} to describe the major states of matter by a single model we introduce the following Higgs-like potential with a metastable vacuum which leads to spontaneous symmetry breaking
\begin{equation}
H_{int}=\sum_{0\leq\hat{\w}_q\leq\omega_{\rm D}}\left[-\frac{\sigma}{2} |Q_{q}^{\a}|^4+\frac{\theta}{6}|Q_{q}^{\a}|^6\right]
\label{Hint}
\end{equation}
Here, $q$ is a multiindex $\{q_1,q_2,q_3\}$, $\Omega_{\rm D}$ is the Debye frequency, the parameter $\mu$ that takes values 1 or 0,  $\sigma,\theta\in\mathbb{R}^+$ are some real non-negative couplings \cite{sci1}. The collective canonical coordinates $\P^\a_k$ and $Q_k^\a$ are introduced as follows
\begin{equation}
\begin{aligned}
Q_q^\a &=  \sqrt{m} \sum_{j=1}^{N} e^{{\texttt i}L (j\cdot q) }x^\a_j,\\
\P^\a_q &= \dot{Q}_q^\a.
\end{aligned}
\end{equation}
Here, $x_j^\a$  is space coordinate of an atom of a lattice sitting in a vertex labelled by the multiindex $j=(j_1,j_2,j_3)$ whose components run $j_i\in (1,N_i)$. Hence the total number of atoms is $N=N_1N_2N_3$. The row $N=(N_1,N_2,N_3)$ gives the maximal size of the lattice in each direction. In addition, $L^3$ is the size of each lattice cell, ${\texttt i}$ is the imaginary unit  (${\texttt i}^2=-1$) and $m$ is the mass of an atom in the lattice. To keep the coordinates  $x^\a_j$  real, one imposes the condition $Q_q^\a=Q_{-q}^\a{}^*$, where star denotes the complex conjugation. 

The above construction is the discrete version of the inverse Fourier transform
\begin{equation}
\hat{\f}_{\vec{q}}=\int d^3 y e^{-i \vec{y}\cdot \vec{q}} \f (\vec{y}).
\end{equation}
However, we avoid using the vector notation for multiindex labelling a lattice vertex so as to not to confuse it with  a space vector labelled by small Greek indices.

Note, that the model is defined for the canonical collective coordinates $\Pi^\a_q$ and $Q_q^\a$ that describe dynamics of phonons in the reciprocal space. Hence, although we have started with a lattice of finite spacing  $L$ there is still an $SO(3)$ symmetry associated with the collective excitations acting on the small Greek indices
\begin{equation}
\begin{aligned}
Q_q^\a & \rightarrow R^\a{}_\b Q_q^\b,\\
\P_q^\a & \rightarrow R^\a{}_\b \P_q^\b.
\end{aligned}
\end{equation}
The Hamiltonian is apparently invariant under these transformations for any $R\in SO(3)$. Obviously, this transformation acts only on the phonon fields and does not transform the lattice itself.

\begin{figure}
	\centering
\includegraphics[scale=0.09]{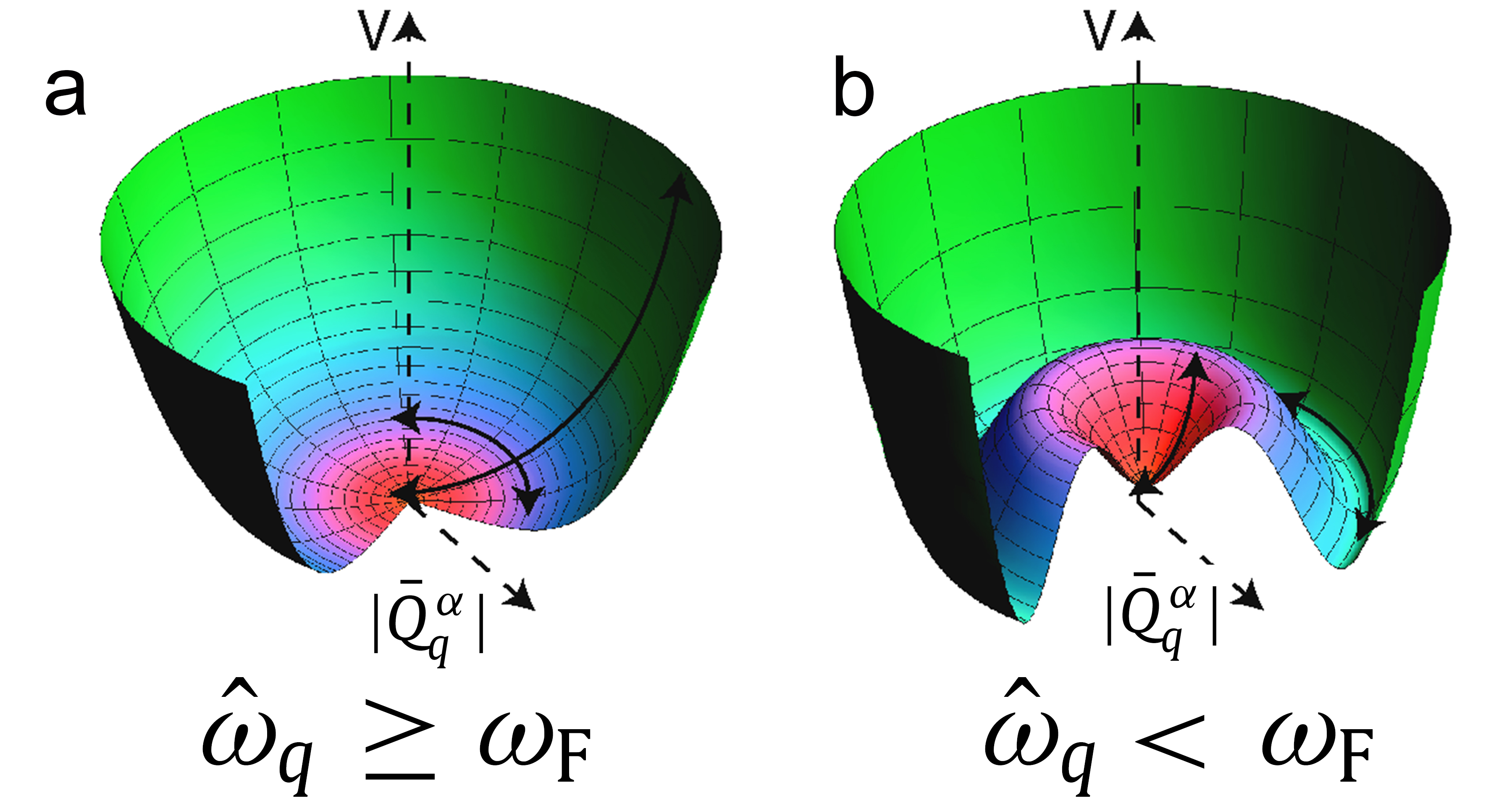}
\caption{(a). Modes with $\hat{\w}_q\geq\w_{\rm F}$ propagate over the trivial vacuum $|\bar{Q}_q^\a|_{0}=0$, where $SO(3)\to SO(3)$. (b) Modes with $\hat{\w}_q<\w_{\rm F}$ propagate over the vacuum $|\bar{Q}^\a_q|$ with broken symmetry, where $SO(3)\to SO(2)$.}
	\label{fig2}
\end{figure}

The kinetic energy is minimal at the static configurations $\bar{\Pi}_q^\a=0$ and minima of the potential
\begin{equation}
V[Q_q^\a]=\sum_{0\leq\hat{\w}_q\leq\omega_{\rm D}}\left[\frac{\mu\hat{\w}_q^2}{2} |Q_q^\a|^2-\frac{\sigma}{2} |Q_{q}^\a|^{4}+\frac{\theta}{6}|Q_{q}^\a|^6\right]
\end{equation}
can be found in the standard way by varying along the coordinate $\frac{\delta V[Q_q^\a]}{\delta Q^\beta_q}=0$. In general, there are six solutions to a six order polynomial. However in this case two roots coincides, and two are negative and hence should be dropped.  In total, keeping only non-negative solutions we have
\begin{equation}
\label{vac}
\begin{aligned}
|\bar{Q}_q^\a|& =\left(\frac{\sigma}{\theta}+ \sqrt{\frac{\omega_{\rm F}^2-\hat{\w}_q^2}{\theta}}\right)^{1/2},\\
|\bar{Q}_q^\a|_{0}&=0,\\
\omega_{\rm F}& = \sqrt{\frac{\sigma^2}{\theta}}.
\end{aligned}
\end{equation}
where $\omega_{\rm F}$ plays the role of the Frenkel frequency (see Fig. (\ref{fig2})).  The square root in the non-trivial solutions $|\bar{Q}_q^\a|$ implies that $\w_{\rm F}$ cuts the spectrum of modes and is related to the viscosity and infinite-frequency shear modulus of a liquid. Indeed, using the Maxwell relation $\tau=\frac{\eta}{G_{\infty}}$ the Frenkel frequency can be represented as
\begin{equation}
\omega_{\rm F}=\sqrt{\frac{\sigma^2}{\theta}}=\frac{2\pi}{\tau}=\frac{2\pi G_{\infty}}{\eta},
\label{ff}
\end{equation}
where $\eta$ is viscosity, G$_{\infty}$ is the infinite-frequency shear modulus and $\tau$ is the average time between two consecutive atomic jumps at one point in space. This equation is noteworthy since it relates parameters of the potential to measurable variables.

Oscillation frequency of excitations $\hat{\w}_q$ in a solid body is bound from above by the Debye frequency $\w_{\rm D}$. Hence, in the most general case one has $\w_{\rm F} \leq \w_{\rm D}$. This implies very different behaviour of modes with the oscillation frequency $\hat{\w}_q$ above and below the Frenkel frequency $\w_{\rm F}$. Indeed, for modes with $\hat{\w}_q \geq \w_{\rm F}$ one has to keep the solution $|\bar{Q}_q^\a|_{0}$, and hence phonons propagate over the trivial vacuum with unbroken $SO(3)$ symmetry. As we will see, this corresponds to keeping all three modes, longitudinal and two transverse, in the spectrum.

In contrast, the trivial vacuum appears to be not the true vacuum for modes with $\hat{\w}_q\leq\w_{\rm F}$. These are allowed to propagate over the vacuum $|\bar{Q}_q^\a|$ with broken symmetry. This corresponds to removing transverse modes from the spectrum. Hence, we see that symmetry breaking occurs when the frequency $\hat{\w}_q$ crosses the bound $\w_{\rm F}$.

Hamiltonian over the vacuum $|\bar{Q}^a_q|_0$ can be obtained in a trivial way and basically has the same form as the initial one. Consider now excitations of the phonon field around the ground state $\bar{Q}_q^\a$ 
\begin{equation}
Q^\a_q=\bar{Q}_q^\a+\varphi_q^\a.
\label{vacuum}
\end{equation}
 For a particularly chosen vacuum $\bar{Q}_q^\a=\delta^\a_1|\bar{Q}_q|$ we obtain the following Hamiltonian
\begin{equation}
\label{Ham}
\begin{aligned}
H[\varphi^\a_q]& =\frac{1}{2}\sum_{0\leq\hat{\w}_q^{l,t,t}\leq\omega_{\rm D}}\pi_q^\a\pi_{-q}^\a+\sum_{0\leq\hat{\w}_q^l\leq\omega_{\rm D}}\left[\frac{\m \w_q^2}{2}\varphi^1_q\varphi^1_{-q}\right]+\\
&+ \sum_{\omega_{\rm F}\leq\hat{\w}_q^{t,t}\leq\omega_{\rm D}}\left[\frac{\mu\w_q^2}{2}(\varphi^2_q\varphi^2_{-q}+\varphi^3_q\varphi^3_{-q})\right]\\
&+V_{int}[\f_q^\a]+V_0,
\end{aligned}
\end{equation}
where $\varphi_q^{2,3}$ we call transverse modes and the mode $\varphi^1_q$ we call longitudinal since it becomes naturally separated. The corresponding canonical momenta are denoted by $\p_q^\a=\dot{\f}_q^\a$. All terms of the order higher than two in fields are collected into the interaction potential $V_{int}$. A constant term $V_0$ remains that does not affect the dynamics and can be dropped. In addition we define oscillation frequency $\w_q$ as follows
\begin{equation}
\w_q^2=
\left\{
\begin{aligned}
& 4\left(\w_{\rm F}^2-\hat{\w}_q^2\right)+ 4\w_{\rm F}\sqrt{\w_{\rm F}^2-\hat{\w}_q^2}>0,& \hat{\w}_q < \w_{\rm F}, \\
& \hat{\w}_q^2, & \hat{\w}_q\geq \w_{\rm F}.
\end{aligned}
\right.
\end{equation}
It is important to mention that $\W_q=\W_q(\w_q)$ is continuous at the point $\w_q^l=\hat{\w}_q=\w_{\rm F}$ and hence the longitudinal mode does not fill the bound between low- and high-frequency regimes: $0\leq\w_{q}^{l}<\w_{\rm F}$ and $w_{\rm F}\leq\w_{q}^{l}\leq\w_{\rm D}$, respectively. For the transverse modes $\f_q^{2,3}$ the situation is drastically different. Due to the Goldstone theorem there are no terms quadratic in $\f_q^{2,3}$ in the region $\w_q^{t,t}<\w_{\rm F}$ where the symmetry is broken. 

In quantum field theory (QFT) this situation corresponds to having massless excitations in the spectrum. From the point of view of condensed matter physics this implies a dynamical cut-off of transverse modes. In other words, there are no transverse modes with $0\leq\w_q^{t,t}<\w_{\rm F}$. Such behaviour allows reproducing energy spectrum of a gas,  a liquid and a solid in a single framework. Referring the reader to \cite{sci1} for details we repeat here the table with results for completeness.
\begin{table}[ht]
\label{tab}
\caption{States of Matter. {\bf Ideal Gas}: no collective excitations; {\bf Fluid below Frenkel line}: only longitudinal excitations $0\leq\omega_{q}^{l}\leq\omega_{\rm D}$; {\bf Liquid above Frenkel line}: both longitudinal ($0\leq\omega_{q}^{l}\leq\omega_{\rm D}$) and transverse ($\omega_{\rm F}\leq\omega_{q}^{t,t}\leq\omega_{\rm D}$) modes; and {\bf Solid}: all modes are supported ($0\leq\omega_{q}^{l,t,t}\leq\omega_{\rm D}$).}
\centering
\begin{tabular}{p{2cm}p{3.5cm}p{1.5cm}}
\hline
\hline \\[-0.2cm]
Phase & Coupling constants & Normal modes \\[0.5ex]
\hline \\[-1.5ex]
{\begin{tabular}{ll}Ideal \\ Gas\end{tabular}}  &  $ \begin{aligned}
			& \m=0,  & &  \s\to 0, \\
            & \q\to0, & & \displaystyle \sqrt{\fr{\s^2}{\q}}=\w_{\rm F}\to 0
			 \end{aligned} $& \begin{tabular}{l}$|\bar{Q}|=0$\\ no modes \end{tabular} \\[0.6cm]
\hline \\[-0.2cm]
 {\begin{tabular}{ll}Fluid below\\ Frenkel line\end{tabular}} \  & $ \begin{aligned}
  			& \m=1,  & &  \s\neq 0, \\
              & \q\neq0, & & \displaystyle \sqrt{\fr{\s^2}{\q}}=\w_{\rm F}=\w_{\rm D}
  			 \end{aligned} $    & $\begin{aligned} & |\bar{Q}|\neq 0, \\
  			\ \  & \f_{q}^{1} \end{aligned}$ \\[0.6cm]			
\hline \\[-0.2cm]
 {\begin{tabular}{ll}Liquid above\\ Frenkel line\end{tabular}} \    & $ \begin{aligned}
 			& \m=1,  & &  \s\neq 0, \\
             & \q\neq0, & & \displaystyle \sqrt{\fr{\s^2}{\q}}=\w_{\rm F}\neq 0
 			 \end{aligned} $     & $ \begin{aligned} & |\bar{Q}|\neq 0, \\ \ &\f_{q}^{1,2,3} \end{aligned}$  \\[0.6cm]
\hline \\[-0.2cm]
{\begin{tabular}{ll}Solid\end{tabular}}   & $ \begin{aligned}
 			& \m=1,  & &  \s\ll\q, \\
             & \q\neq0, & & \displaystyle \sqrt{\fr{\s^2}{\q}}=\w_{\rm F}\to 0
 			 \end{aligned} $    & $ \begin{aligned} & |\bar{Q}|\neq 0, \\ &\f_{q}^{1,2,3} \end{aligned}$ \\[0.6cm]
\hline
\hline \\[-0.2cm]
\end{tabular}

\end{table}

We note, that thermodynamic limits can be derived in accordance with the equipartition theorem on the classical basis only while the present formalism covers the quantum limit, classical limit for solids (the Dulong-Petit law), {\it rigid liquid} regime, the Frenkel line thermodynamic limit, {\it non-rigid fluid} regime and the ideal gas thermodynamic limit. In the vicinity of the melting point the heat capacity at constant volume per particle $c_{V}$ of a liquid customarily goes beyond the Dulong-Petit law ($c_V=3k_{\rm B}$) and our theory covers it too by incorporating anharmonic effects \cite{bolprb}. All regimes of our theory are summarized in Table I.
\section{Statistical mechanics of aggregation states}
\subsection{Debye approximation}
The phonon density of states D($\omega$) describes the number of phonon modes of a selected frequency $\omega$ in a given frequency interval and can be presented as
\begin{equation}
D(\omega)=\frac{\omega^2}{q^3_{\rm D}}\left[\left(\frac{q^2}{\omega^2}\frac{\partial q}{\partial\omega}\right)_{\rm L}+2\left(\frac{q^2}{\omega^2}\frac{\partial q}{\partial\omega}\right)_{\rm T}\right]
\end{equation}
where L and T stand for the longitudinal and transverse phonon polarizations, respectively \cite{llandau}. 

In the Einstein approximation the phonon density of states is expressed via the Dirac $\delta$-function \cite{aeinstein}, while in the Debye approximation the phonon density of states $D(\omega)\sim\omega^2$ \cite{pdebye}. Here, we approximate the normal vibrations with elastic vibrations of an isotropic continuous body and assume that the number of vibrational modes $D(\omega)d\omega$ having angular frequencies between $\omega$ and $\omega+d\omega$ is given by 
\begin{equation}
\label{vac}
\begin{aligned}
D(\omega)& =\frac{V}{2\pi^2}\left(\frac{1}{c_{l}^{3}}+\frac{2}{c_{t}^{3}}\right)\equiv\frac{9 N}{\omega^3_{\rm D}}\omega^2 \ & (\omega\leq\omega_{\rm D}), \\
D(\omega)&=0 \ & (\omega>\omega_{\rm D}).
\end{aligned}
\end{equation}
where $c_{l}$ and $c_{t}$ are the velocities of longitudinal and transverse waves, respectively. The Debye frequency is determined by
\begin{equation}
\int_{0}^{\omega_{\rm D}}g(\omega)d\omega\equiv 3N
\end{equation}
where $N$ is the number of atoms \cite{pdebye}. 

Heat capacity is considered to be one of the most important properties of matter because it holds information about system's degrees of freedom as well as the regime in which the system operates, classical or quantum. Therefore, calculating the partition function $Z$ is the central problem of statistical mechanics. Once the system's partition function has been calculated, its Helmholtz free energy $F$ follows: $F=-k_{\rm B}T\ln{Z}$. The internal energy $E$ is given by: $E=-T^2\left[\frac{\partial}{\partial T}\left(\frac{F}{T} \right)\right]_{\rm V,N}$. Accordingly, the heat capacity at constant volume per atom is given by: $c_V=\frac{1}{N}\left(\frac{\partial E}{\partial T}\right)_{\rm V}$.

In general, the velocity of sound propagation depends on the phonon polarization ($c_l\neq c_t$). Therefore, we employ the Debye approximation 
\begin{equation}
c_l=c_t=c \ \Longrightarrow \\
 \w_q=cq
\label{da}
\end{equation}
to derive the energy spectra of aggregation states from the effective Hamiltonian (\ref{Ham}). Dropping the arbitrary constant term $V_0$ and all terms of higher degree in the fields,  the effective Hamiltonian takes the following form
\begin{equation}
\label{EHam}
\begin{aligned}
H[\varphi^\a_q]& =\frac{1}{2}\sum_{0\leq\hat{\w}_q^{l,t,t}\leq\omega_{\rm D}}\pi_q^\a\pi_{-q}^\a+\sum_{0\leq\hat{\w}_q^l\leq\omega_{\rm D}}\left[\frac{\m \w_q^2}{2}\varphi^1_q\varphi^1_{-q}\right]+\\
&+ \sum_{\omega_{\rm F}\leq\hat{\w}_q^{t,t}\leq\omega_{\rm D}}\left[\frac{\mu\w_q^2}{2}(\varphi^2_q\varphi^2_{-q}+\varphi^3_q\varphi^3_{-q})\right]
\end{aligned}
\end{equation}
and
\begin{equation}
\omega_{\rm F}(T)=\sqrt{\frac{\sigma^2}{\theta}}=\frac{2\pi}{\tau(T)}=\frac{2\pi G_{\infty}}{\eta(T)}.
\label{ff}
\end{equation}

\subsection{Thermodynamic limits: regimes and predictions of the theory}
In this section we derive the energy spectra of basic states of matter from the effective Hamiltonian (\ref{EHam}). Each energy spectrum has an exact analytical expression with no free fitting and adjustable  parameters. Starting with the effective Hamiltonian (\ref{EHam}) and applying the Debye approximation we construct a generalization of the continuum theory of solids by Debye, the phonon theory of liquids and the supercritical state, and also cover thermodynamic limits: the quantum limit ($c_V\propto T^{3}$), the Dulong-Petit law ($c_V=3k_{\rm B}$), the Frenkel line limit ($c_V=2k_{\rm B}$), and the ideal gas thermodynamic limit ($c_V=\frac{3}{2}k_{\rm B}$).
\subsubsection{Rigid liquids, c$_V$: $3$k$_{\rm B}\longrightarrow$ $2$k$_{\rm B}$}
\begin{figure}
	\centering
\includegraphics[scale=0.063]{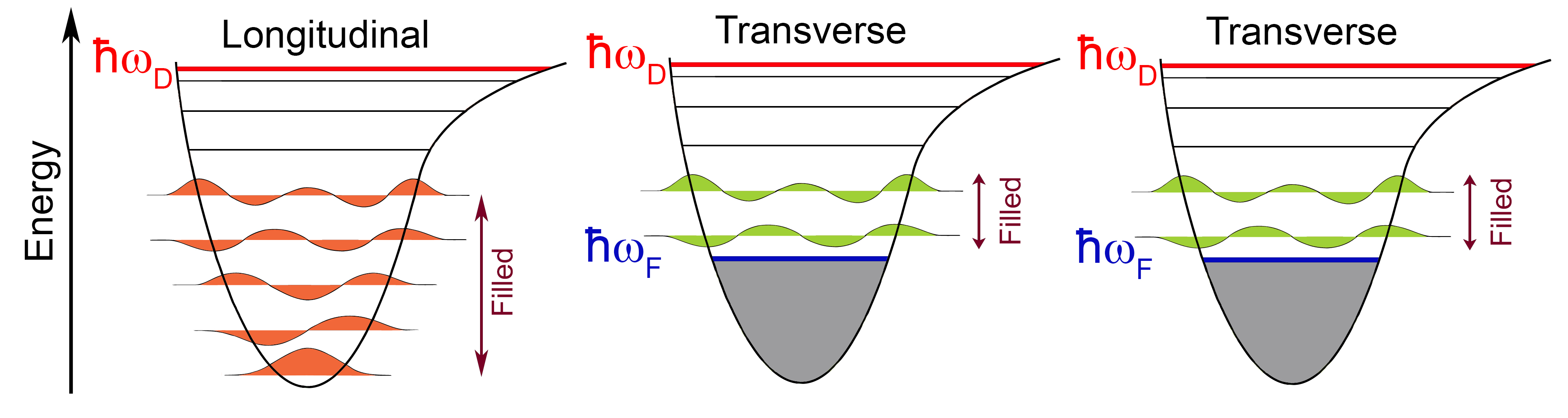}
\caption{Vibrational potential energy distribution in longitudinal and transverse phonon polarisations in the {\it rigid liquid}  thermodynamic regime. In the longitudinal direction phonons can be excited similarly as in disordered solids. In the transverse directions the vibrational energy states start filling from the lower bound defined by the Frenkel energy $\hslash\omega_{\rm F}$. The energy gap in transverse phonon excitations emerges naturally due to symmetry breaking in phonon excitations (see Eq. \ref{Hint}).}
	\label{erigid}
\end{figure}

Here, we set parameters in the effective Hamiltonian (\ref{EHam}) as
\begin{equation}
\mu=1, \ \sigma\neq 0, \ \theta\neq 0, \ \frac{\sigma^2}{\theta}\neq 0
\end{equation}
where normal modes (1 longitudinal phonon mode: $0\leq\omega_{q}^{l}\leq\omega_{\rm D}$) and (2 transverse phonon modes: $\omega_{\rm F}\leq\omega_{q}^{t}\leq\omega_{\rm D}$, see Fig. (\ref{erigid}))
\begin{equation}
|\bar{Q}|\neq 0, \ \phi_{q}^{1,2,3} \left(\omega_{q}^{t,t}\geq\omega_{\rm F}\right)
\end{equation}
We calculate the partition function for a harmonic oscillator having an angular frequency $\omega_{i}$ 
\begin{equation}
Z_{i}=\sum_{0}^{\infty}e^{\frac{-(n+\frac{1}{2})\hslash\omega_{i}}{k_{\rm B}T}}=\left[2\sinh{\frac{\hslash\omega_{i}}{2k_{\rm B}T}} \right]^{-1}
\label{partition}
\end{equation}
The partition function for a system of identical harmonic oscillators having various frequencies is given by 
\begin{equation}
Z=\prod Z_{i}
\end{equation}
where oscillators numbered by $i$. Accordingly, the Helmholtz free energy is given by
\begin{equation}
F=E_{0}+k_{\rm B}T\sum_{i}\ln{\left(1-e^{-\frac{\hslash\omega_{i}}{k_{\rm B}T}} \right)}
\end{equation}
where $E_0$ is the energy of zero-point vibrations. We take into account the effect of thermal expansion which implies
\begin{equation}
\frac{\partial\omega}{\partial T}\neq 0
\end{equation}
The internal energy is given by
\begin{equation}
E=E_0+\hslash\sum_{i}\frac{\omega_{i}-T\frac{d\omega_{i}}{dT}}{e^{\frac{\hslash\omega_{i}}{k_{\rm B}T}}-1}
\label{energy}
\end{equation}
which is contrary to the harmonic case. Here, we employ quasi-harmonic Gr\"uneisen approximation (see $Anharmonicity$ section for detail) which leads to $\frac{\partial\omega}{\partial T}=-\frac{\alpha\omega}{2}$, where  $\alpha$ is the coefficient of thermal expansion. Inserting it into Eq. (\ref{energy}) gives
\begin{equation}
E=E_0+\left(1+\frac{\alpha T}{2}\right)\sum_{i}\frac{\hslash\omega_{i}}{e^{\frac{\hslash\omega_{i}}{k_{\rm B}T}}-1}
\end{equation}

Hence, the internal energy of {\it rigid liquid} can be represented in the final form
\begin{equation}
E=NT\left(1+\frac{\alpha T}{2}\right)\left(3D\left(\frac{\hslash\omega_{\rm D}}{k_{\rm B}T}\right)- \left(\frac{\omega_{\rm F}}{\omega_{\rm D}}\right)^{3}D\left(\frac{\hslash\omega_{\rm F}}{k_{\rm B}T}\right)\right)
\label{E}
\end{equation}
where
\begin{equation}
D(x)=\frac{3}{x^3}\int_{0}^{x}\frac{\varsigma^{3}d\varsigma}{e^{\varsigma}-1}
\label{D}
\end{equation}
is the Debye function. Eq. (\ref{E}) spans both quantum and classical limits which we will discuss in detail below. Heat capacity at constant volume per atom $c_V$ can readily be calculated from Eq. (\ref{E}): $c_V=\frac{1}{N}\left(\frac{\partial E}{\partial T}\right)_{\rm V}$. In the {\it rigid liquid} regime $c_V$ drops down from about $3k_{\rm B}$ (Dulong-Petit law value) to approximately $2k_{\rm B}$ (The Frenkel line thermodynamic limit).

\subsubsection{Dulong-Petit thermodynamic limit ( c$_V$ = $3$k$_{\rm B}) \ and \ beyond$}
From Eq. (\ref{E}) we can immediately derive the empirical results for $c_V$ discovered by Dulong and Petit in the early 1800s. For $x\longrightarrow 0$ it gives $D(x)=1$. Therefore, at elevated temperatures when $T\longrightarrow \infty$ it leads to $D\left(\frac{\hslash\omega_{\rm D}}{k_{\rm B}T}\right)\longrightarrow 1$. In solid phase $\omega_{\rm F}=0$ and when $T\longrightarrow \infty$ it yields the same result for the Debye function $D\left(\frac{\hslash\omega_{\rm F}}{k_{\rm B}T}\right)\longrightarrow 1$. Whence, we obtain internal energy and heat capacity beyond the Dulong-Petit thermodynamic limit from Eq. (\ref{E}) (or from Eq. 12, \cite{bolprb})
\begin{equation}
\begin{aligned}
E &=3Nk_{\rm B}T\left(1+\frac{\alpha T}{2}\right)  \\
c_V &=\frac{C_V}{N}=3k_{\rm B}(1+\alpha T)
\end{aligned}
\label{beyond}
\end{equation}
In the framework of the harmonic model we yield the Dulong-Petit law from Eq. (\ref{E})  
\begin{equation}
C_V=Nk_{\rm B}\frac{\partial }{\partial T}\cdot T(1+0)(3\cdot 1-0)=3Nk_{\rm B}
\end{equation}
or equivalently  putting $\alpha=0$ in Eq. (\ref{beyond}) we obtain 
\begin{equation}
c_V=3k_{\rm B}
\end{equation}

\subsubsection{The Frenkel line thermodynamic limit: c$_V$ = $2$k$_{\rm B}$}
The Frenkel line is the new thermodynamic boundary on the P-T diagram which was recently experimentally evidenced in a deeply supercritical sample through diffraction measurements in a diamond anvil cell \cite{bolf}. 

Approaching the Frenkel line corresponds to the qualitative change of atomic dynamics in a liquid. In  rigid/ compressed liquids, atomic motion has a vibrational component of motion about equilibrium positions and diffusive component of motion between adjacent equilibrium positions. As the temperature grows, an atom spends less time oscillating and more time diffusing. Hence, the oscillating component of motion eliminates \cite{brazhkintoday}. That disappearance of oscillatory component of motion corresponds to crossing the Frenkel line: the transition of the substance from the liquid dynamics to the gas dynamics \cite{brazhkinprl}. This crossover takes place when liquid relaxation time $\tau$ ($\tau$ is liquid relaxation time, the average time between consecutive atomic jumps at one point in space \cite{ffrenkel}) approaches its minimal value, the $\tau_{\rm D}$ Debye vibration period. At relatively high temperatures  we may employ the limit $T\longrightarrow \infty$ which yields $D\left(\frac{\hslash\omega_{\rm D}}{k_{\rm B}T}\right)\longrightarrow 1$ and $D\left(\frac{\hslash\omega_{\rm F}}{k_{\rm B}T}\right)\longrightarrow 1$. In the rigid liquid regime the Frenkel frequency $\omega_{\rm F}\neq 0$ and can be presented through the relaxation time or through the shear modulus and the viscosity $\omega_{\rm F}=\frac{2\pi}{\tau}=\frac{2\pi G_{\infty}}{\eta}$ (see Eq. (\ref{ff})). The viscosity behaviour of most compressed liquids can be described by Vogel-Fulcher-Tammann (VFT) law \cite{bolprb}
\begin{equation}
\eta=\eta_{0}\exp{\left(\frac{A}{T- T_{0}} \right)}
\end{equation}
As viscosity approaches its minimal value and the Frenkel frequency $\omega_{\rm F}$ becomes comparable to the Debye frequency $\omega_{\rm D}$ ($\omega_{\rm F}\xrightarrow{T}\omega_{\rm D}$, see Fig. (\ref{enonrigid}))  Eq. (\ref{E}) leads to the thermodynamic limit for internal energy 
\begin{equation}
E=NT\left(1+\frac{\alpha T}{2}\right)\left(3\cdot 1-\left( \frac{\omega_{\rm F}\xrightarrow{T}\omega_{\rm D}}{\omega_{\rm D}}\right)^{3}\cdot 1\right)
\end{equation}
The Frenkel line thermodynamic limit implies that heat capacity at constant volume 
\begin{equation}
c_V=\frac{1}{N}\left(\frac{\partial E}{\partial T}\right)_{\rm V}=2k_{\rm B}, \ when \ \ \alpha=0
\end{equation}
whence, the Frenkel line is a smooth and continuous boundary which originates in the neighborhood of the critical point and extends into the supercritical region up to arbitrarily high pressures and temperatures. Strictly speaking, however, $c_V\approx 2k_{\rm B}$; not exactly $c_V=2k_{\rm B}$ due to anharmonic effects ($\alpha\neq 0$) \cite{bolnature}. Therefore, $c_V=2k_{\rm B}$ is the thermodynamic limit but the boundary (the Frenkel line) is smeared out in its vicinity. 
\subsubsection{Non-rigid fluids, c$_V$: $2$k$_{\rm B}\longrightarrow$ $\frac{3}{2}$k$_{\rm B}$}
\begin{figure}
	\centering
\includegraphics[scale=0.063]{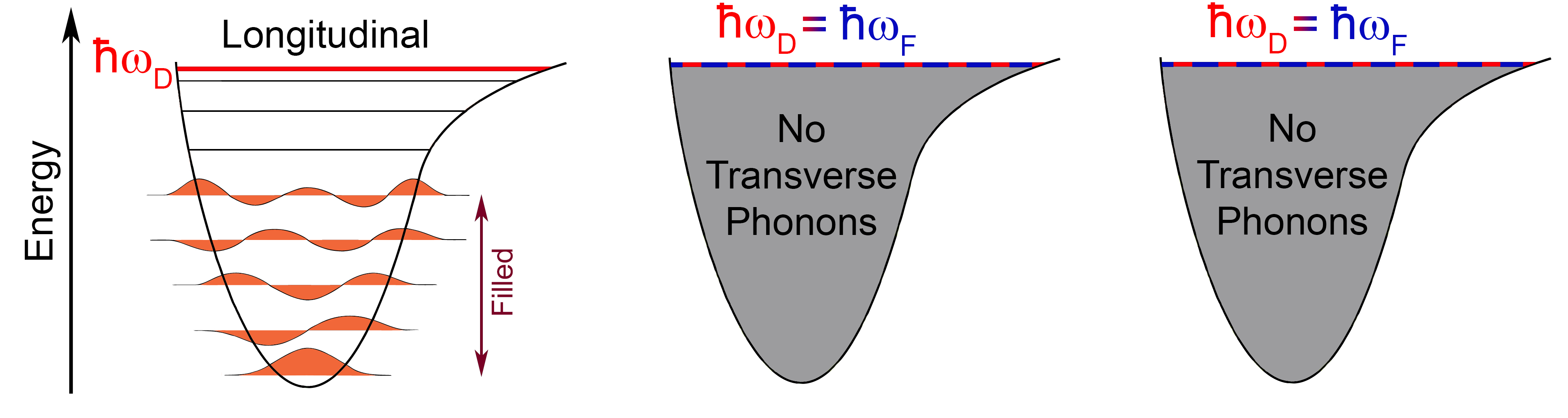}
\caption{Vibrational potential energy distribution in longitudinal and transverse phonon polarisations in the {\it non-rigid fluid}  thermodynamic regime. In the longitudinal direction phonons can be excited similarly as in disordered solids and $rigid$ liquids. In the transverse directions phonons cannot be excited due to $\hslash\omega_{\rm F}=\hslash\omega_{\rm D}$.}
	\label{enonrigid}
\end{figure}

In the case of {\it non-rigid fluid} regime we set parameters in the effective Hamiltonian (\ref{EHam}) as
\begin{equation}
\mu=1, \ \sigma\neq 0, \ \theta\neq 0, \ \frac{ \sigma^2}{\theta}=\omega_{\rm F}=\omega_{\rm D}
\end{equation}
and only one normal mode exists in the {\it non-rigid fluid} regime (1 longitudinal phonon mode: $0\leq\omega_{q}^{l}\leq\omega_{\rm D}$, see Fig. (\ref{enonrigid}))
\begin{equation}
|\bar{Q}|\neq 0, \ \phi_{q}^{1}
\end{equation}
Hence, the effective Hamiltonian ( see Eq. \ref{EHam}) becomes
\begin{equation}
\label{NR}
\begin{aligned}
H[\varphi_q] =\frac{1}{2}\sum_{0\leq\omega_q^{l,t,t}\leq\omega_{\rm D}}[\pi_q^{1}\pi_{-q}^{1}+\pi_q^{2}\pi_{-q}^{2}+\pi_q^{3}\pi_{-q}^{3}]+\\
\frac{1}{2}\sum_{0\leq\omega_q^l\leq\omega_{\rm D}}\left[\omega_q^2\varphi^1_q\varphi^1_{-q}\right]
\end{aligned}
\end{equation}

When the Frenkel line crossed the solid-like oscillating component of motion \cite{sci1}, two transverse phonon modes \cite{bolsrph} and medium-range order disappear \cite{bolstr1,bolstr2}. All that remain is one longitudinal phonon mode, gas-like ballistic motion and short-range order pair correlations. That disappearance, a qualitative change in particle thermodynamics, dynamics and structure corresponds to the thermodynamic \cite{bolnature,sci1,bolsrph}, dynamic \cite{sci1,bolsrph,brazhkinprl} and structural \cite{bolstr1,bolstr2} crossovers at the Frenkel line, respectively. 

As temperature $T$ increases in the $non$-$rigid$ $fluid$ regime and kinetic energy rises, the mean free path $l$, the average distance between particle collisions, increases. At the Frenkel line where the ballistic regime starts, $l$ is comparable to interatomic separation $a$. In the limit of high temperature where the particle's kinetic energy is much larger than potential energy, $l$ tends to infinity as in the non-interacting ideal gas limit. Our proposal is that $l$ determines the shortest wavelength of the longitudinal phonon mode that exists in the system, $\lambda$, because below this length the motion is purely ballistic and therefore can not be oscillatory, $\lambda=l$. On the other hand, the longitudinal phonon excitations with larger wavelength are supported and they represent the excitations existing in the supercritical system \cite{bolnature}. 

Then, the energy of non-rigid gas-like supercritical fluid becomes
\begin{equation}
E=\frac{3}{2}Nk_{\rm B}T+\left(1+\frac{\alpha T}{2}\right)\frac{1}{2}Nk_{\rm B}T\frac{a^3}{\lambda^3}
\label{fen}
\end{equation}
At the Frenkel line $\lambda=a$ in harmonic approximation ($\alpha=0$) Eq. (\ref{fen}) gives $E=2Nk_{\rm B}T$ and $c_{V}=\frac{C_{V}}{N}=2k_{\rm B}$. This corresponds to the crossover of heat capacity at constant volume  from the {\it rigid liquid} regime \cite{bolsrph} to the {\it non-rigid fluid} regime \cite{bolnature}, respectively. When $\lambda$ rises as $T$ increases in the {\it non-rigid fluid} regime equation (\ref{fen}) predicts  
\begin{equation}
c_V\xrightarrow{T}\frac{3}{2}k_{\rm B}
\end{equation}
where c$_{V}=\frac{3}{2}$k$_{\rm B}$ is the ideal gas thermodynamic limit.

\subsubsection{ Low-$T$ quantum limit: c$_V\sim$T${^3}$}
\begin{figure}
	\centering
\includegraphics[scale=0.063]{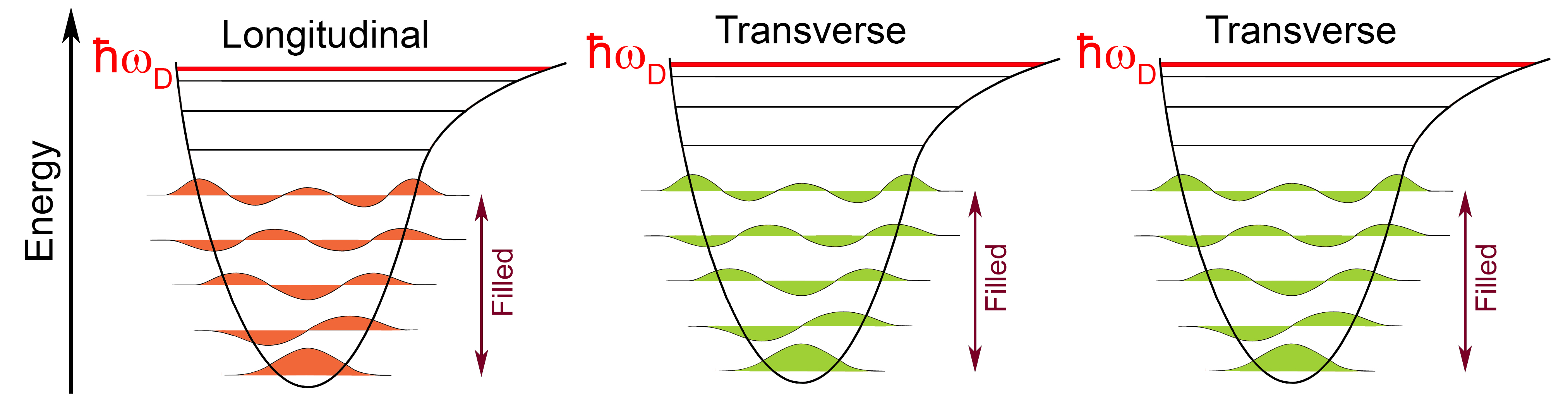}
\caption{Vibrational potential energy distribution in longitudinal and transverse phonon polarisations in the {\it solid}  thermodynamic regime. Phonons can be excited in any direction of wave-packet propagation.}
	\label{esolid}
\end{figure}

There is no described symmetry breaking in phonon interactions \cite{sci1}.  In case of solids and in the quantum regime in particular we set parameters in the effective Hamiltonian (\ref{EHam}) as
\begin{equation}
\mu=1, \ \sigma\ll \theta, \ \theta\neq 0, \ \frac{ \sigma^2}{\theta}\rightarrow 0
\end{equation}
all modes are supported (1 longitudinal and 2 transverse $0\leq\omega_{q}^{l,t,t}\leq\omega_{\rm D}$, see Fig. (\ref{esolid}))
\begin{equation}
|\bar{Q}|\neq 0, \ \phi_{q}^{1,2,3}
\end{equation}
Hence, the effective Hamiltonian (see Eq. \ref{EHam}) yields
\begin{equation}
\label{EHam1}
\begin{aligned}
H[\varphi_q] =\frac{1}{2}\sum_{0\leq\omega_q^{l,t,t}\leq\omega_{\rm D}}[\pi_q^{1}\pi_{-q}^{1}+\pi_q^{2}\pi_{-q}^{2}+\pi_q^{3}\pi_{-q}^{3}]+\\
\frac{1}{2}\sum_{0\leq\omega_q^{l,t,t}\leq\omega_{\rm D}}\left[\omega_q^2(\varphi^1_q\varphi^1_{-q}+\varphi^2_q\varphi^2_{-q}+\varphi^3_q\varphi^3_{-q})\right]
\end{aligned}
\end{equation}
and the Frenkel frequency 
\begin{equation}
\omega_{\rm F}=\frac{2\pi}{\tau\rightarrow\infty}=\frac{2\pi G_{\infty}}{\eta\rightarrow\infty}=0
\end{equation}
reflecting the fact that solids are not able to flow (viscosity $\eta\rightarrow\infty$).
Then, it becomes trivial to derive the internal energy and the heat capacity from the partition function ( see Eq. (\ref{partition})) following the standard procedure in the framework of the Debye model.

Alternatively, the analytical expression for the internal energy and the heat capacity for solids can be derived from Eq. (\ref{E}) which readily gives
\begin{equation}
E=3Nk_{\rm B}TD\left(\frac{\hslash\omega}{k_{\rm B}T}\right)
\end{equation}
covering the classical limit, the Dulong-Petit law, on one hand $E=3Nk_{\rm B}T\left(D\left(\frac{\hslash\omega}{k_{\rm B}T}\xrightarrow{T\rightarrow\infty}0\right)\rightarrow 1\right)$
\begin{equation}
C_V=3Nk_{\rm B}
\end{equation}
and the quantum limit (low temperature regime)
\begin{equation}
\begin{aligned}
C_V &=3Nk_{\rm B}\frac{3}{\omega_{\rm D}^{3}}\int_{0}^{\omega_{\rm D}}\frac{e^{\frac{\hslash\omega}{k_{\rm B}T}}}{\left(e^{\frac{\hslash\omega}{k_{\rm B}T}}-1\right)^2}\left(\frac{\hslash\omega}{k_{\rm B}T} \right)^{2}D(\omega)d\omega \\
\Rightarrow C_V &\sim T^3
\end{aligned}
\end{equation}
on the other hand.

\begin{table}
\caption{Thermodynamic properties of solid (the Dulong-Petit thermodynamic limit), liquid and gas (ideal gas thermodynamic limit) phases, followed by two regimes in the supercritical region: Rigid Liquid and Non-Rigid Fluid, and the new thermodynamic limit, dubbed the Frenkel line thermodynamic limit.}
\centering
\begin{tabular}{c c c c}
\hline\hline
Phase & $\langle E_{kin}\rangle$, $Nk_{\rm_B}T$ & $\langle E_{pot}\rangle$, $Nk_{\rm_B}T$ & $C_{V}$, $Nk_{\rm_B}$\\ [0.5ex]
\hline
{\begin{tabular}{ll}Solid\end{tabular}}  & $\frac{3}{2}$ & $\frac{3}{2}$ & $3$ \\
{\begin{tabular}{ll}Liquid\end{tabular}} & $\frac{3}{2}$ & $\frac{3}{2}$ $\xrightarrow{T}$ $\frac{1}{2}$ & $3$ $\xrightarrow{T}$ $2$ \\
{\begin{tabular}{ll}Ideal Gas\end{tabular}} & $\frac{3}{2}$ & 0 & $\frac{3}{2}$ \\
\hline\hline
{\begin{tabular}{ll}Rigid \\ Liquid\end{tabular}} & $\frac{3}{2}$ & $\frac{3}{2}$ $\xrightarrow{T}$  $\frac{1}{2}$ & $3$ $\xrightarrow{T}$ $2$ \\
\hline

{\begin{tabular}{ll}Frenkel \\ Line limit\end{tabular}} & $\frac{3}{2}$ & $\frac{1}{2}$ & $2$ \\
\hline
{\begin{tabular}{ll}Non-Rigid \\  Fluid \end{tabular}} &  $\frac{3}{2}$ & $\frac{1}{2}$ $\xrightarrow{T}$ 0 & $2$ $\xrightarrow{T}$ $\frac{3}{2}$  \\
[1ex]
\hline\hline
\end{tabular}

\label{table}
\end{table}

\subsubsection{Ideal gas thermodynamic limit: c$_V$ = $\frac{3}{2}$k$_{\rm B}$}
The ideal gas model depends on the following assumptions: the molecules of the gas are indistinguishable, small hard spheres. All collisions are elastic and all motion is frictionless, i.e. no energy loss in motion or collision. In case of ideal gas thermodynamic limit we set parameters in the effective Hamiltonian (\ref{EHam}) as
\begin{equation}
\mu=0, \ \sigma\rightarrow 0, \ \theta\rightarrow  0, \ \frac{ \sigma^2}{\theta}\rightarrow 0
\end{equation}
and no normal mode exist in the this regime 
\begin{equation}
|\bar{Q}|= 0
\end{equation}
hence
\begin{equation}
\label{ig}
\begin{aligned}
H[\varphi_q] =\frac{1}{2}\sum_{0\leq\omega_q^{l,t,t}\leq\omega_{\rm D}}[\pi_q^{1}\pi_{-q}^{1}+\pi_q^{2}\pi_{-q}^{2}+\pi_q^{3}\pi_{-q}^{3}]
\end{aligned}
\end{equation}
This leaves only the kinetic term in the above equation giving
\begin{equation}
E=k_{\rm B}T^2\frac{\partial \ln{Z}}{\partial T}=\frac{3}{2}Nk_{\rm B}T
\end{equation}
where $Z=$tr$\left( e^{-\frac{H}{k_{\rm B}T}}\right)$. Both longitudinal and transverse phonon modes are non-interacting and massless which corresponds to the ideal gas thermodynamic limit giving
\begin{equation}
c_V=\frac{C_V}{N}\longrightarrow \frac{3}{2}k_{\rm B}
\end{equation}

In Table II we summarise the results from above subsections showing the evolution of kinetic energy, potential energy and heat capacity of a system with temperature variations through solid, liquid and gas phases respectively. We will discuss anharmonic effects in detail below.
\subsection{Anharmonicity}

In the harmonic approximation temperature dependence of frequencies is ignored \cite{bolprb}. It may be a good approximation for phonons in solids at either low temperature or in systems with small thermal expansion coefficient $\alpha$. In liquids, both $rigid$ and $non$-$rigid$ anharmonicity and associated thermal expansion are large and need to be taken into account ($\frac{\partial\omega}{\partial T}\neq 0$) \cite{bolprb}. In this case, applying $E=k_{\rm B}T^2\frac{\partial}{\partial T}\ln Z$ to Eq. (\ref{partition}) gives
\begin{equation}
E=E_{harm}+E_{anharm}
\label{etotal}
\end{equation}
where $E_{harm}$ and $E_{anharm}$ refer to harmonic and anharmonic energy contribution to internal total energy, respectively. The harmonic energy can be represented in the general form as
\begin{equation}
E_{harm}=\sum_{i=1}^{N_{l}}\varepsilon(\omega_{li},T)D(\omega_{li})d\omega+\sum_{i=1}^{N_{t}}\varepsilon(\omega_{ti},T)
D(\omega_{ti})d\omega
\label{eharm}
\end{equation}
where N$_{l}$ and N$_{t}$ is the number of longitudinal and transverse phonon modes, respectively, $0\leq\omega_{l}\leq\omega_{\rm D}$, $\omega_{\rm F}\leq\omega_{t}\leq\omega_{\rm D}$ and $\varepsilon(\omega,T)$ is the mean energy of harmonic oscillators having an angular frequency $\omega$
\begin{equation}
\varepsilon(\omega,T)=\frac{\hslash\omega}{2}\coth{\frac{\hslash\omega}{2k_{\rm B}T}}=\frac{\hslash\omega}{2}+\frac{\hslash\omega}{e^{\frac{\hslash\omega}{k_{\rm B}T}}-1}
\end{equation}
The anharmonic energy can be expressed as
\begin{equation}
E_{anharm}=-k_{\rm B}T^2\left(\sum\limits_{i=1}^{N_{l}}\frac{1}{\omega_{li}}\frac{\partial\omega_{li}}{\partial T}+\sum\limits_{i=1}^{N_t}\frac{1}{\omega_{ti}}\frac{\partial\omega_{ti}}{\partial T}\right)
\label{eanharm}
\end{equation} 
We need to calculate $\frac{\partial\omega_{li}}{\partial T}$ and $\frac{\partial\omega_{ti}}{\partial T}$ in Eq. (\ref{eanharm}) resulting in system softening. It can be calculated in the Gr\"{u}neisen approximation by introducing the Gr\"{u}neisen parameter $\gamma=-\frac{V}{\omega}\left(\frac{\partial\omega_i}{\partial V}\right)_{\rm T}$, where $\omega_i$ are frequencies in harmonic approximation. $\gamma$ features in the phonon pressure $P_{ph}=-\left(\frac{\partial F}{\partial V}\right)_{\rm T}$, where $F$ is the free energy in the harmonic approximation. $F=-k_{\rm B}T\ln Z$ can be calculated from Eq. (\ref{partition}), giving
\begin{equation}
F=k_{\rm B}T\sum\limits_{i=1}^{N_{l}}\ln\frac{\hbar\omega_{li}}{k_{B}T}+k_{\rm B}T\sum\limits_{i=1}^{N_t}\ln\frac{\hbar\omega_{ti}}{k_{\rm B}T}
\end{equation}
Calculating the phonon pressure $P_{ph}=-\left(\frac{\partial F}{\partial V}\right)_T$ and introducing the above Gr\"{u}neisen parameter it leads to $P_{ph}=\frac{\gamma T}{V}(N_{l}+N_{t})$. Then, the bulk modulus due to the negative phonon pressure is $B_{ph}=V\frac{\partial P_{ph}}{\partial V}=-\frac{\gamma T}{V}(N_{l}+N_{t})$, giving $\left(\frac{\partial B_{ph}}{\partial T}\right)_V=-\frac{\gamma}{V}(N_{l}+N_{t})$. We now use the macroscopic definition of $\gamma$ where $\gamma=\frac{V\alpha B}{C_v}$. Here, $B=B_0+B_{ph}$ is the total bulk modulus, $B_0$ is the zero-temperature bulk modulus, $\alpha$ is the coefficient of thermal expansion and $C_V$ is the heat capacity at constant volume. For $C_V$ we use its harmonic representation where $C_V=k_{\rm B}(N_{l}+N_{t})$ from Eq. (\ref{eanharm}), because  $\frac{\partial\omega_{li}}{\partial T}$ and $\frac{\partial\omega_{ti}}{\partial T}$ in Eq. (\ref{eanharm}) already enter as quadratic anharmonic corrections. Therefore, $\left(\frac{\partial B_{ph}}{\partial T}\right)_{\rm V}=-\alpha(B_0+B_{ph})$. For small $\alpha T$, as can be seen from experiments \cite{anderson}, this implies $B\propto -T$.

We note that experimentally, $B$ linearly decreases with $T$ increase at both constant pressure and constant volume \cite{anderson}. The decrease of $B$ with $T$ increase at constant volume is due to the intrinsic anharmonic effects related to the softening of interatomic potential at large vibrational modes. The decrease of $B$ at constant pressure has an additional contribution from the thermal expansion.

Assuming $\omega^2\propto B_0+B_{ph}$ and combining it with $\left(\frac{\partial B_{ph}}{\partial T}\right)_{\rm V}=-\alpha(B_0+B_{ph})$ from above gives $\frac{1}{\omega}\frac{\partial\omega}{\partial T}=-\frac{\alpha}{2}$. Putting it in Eq. (\ref{eanharm}) and combining it with Eq. (\ref{eharm}) we obtain a general analytical expression for internal total energy with account of anharmonic effects
\begin{equation}
E=Nk_{\rm B}T\left(1+\frac{\alpha T}{2}\right)\left(3D\left(\frac{\hslash\omega_{\rm D}}{k_{\rm B}T}\right)- \left(\frac{\omega_{\rm F}}{\omega_{\rm D}}\right)^{3}D\left(\frac{\hslash\omega_{\rm F}}{k_{\rm B}T}\right)\right)
\end{equation}
which covers the Debye theory of solids, the phonon theory of liquids, the Dulong-Petit and the Frenkel line thermodynamics limits, respectively.

\subsection{Phonon propagation and localisation in liquids}
Sound generally travels faster in solids and liquids than in gases. In solids, atoms are tied together with bonds, so they can not vibrate independently. The vibrations take the form of collective modes which propagate through the material. Such propagating lattice vibrations can be considered to be sound waves, and their propagation speed is the speed of sound in the material. In liquids, sound can travel within almost the same frequency range as in solids ($0\leq\omega_q\leq\omega_{\rm D,liquid}\approx\omega_{\rm D,solid}\cdot(\sim 80\%)$) towards wave propagation but in two transverse directions at high frequencies only ($\frac{2\pi G_{\infty}}{\eta}=\omega_{\rm F}\leq\omega_q\leq\omega_{\rm D}$), see Eqs. (\ref{EHam} and \ref{E}).

\subsubsection{"Normal" and "fast sound" propagation in liquids}
In liquids, due to the existence of transverse phonon energy gaps  unusual phenomena associated with sound propagation and localisation of transverse phonon modes  may occur at high and low frequencies regimes, respectively. To understand the nature of propagating and localised phonon modes in liquids we consider the effective Hamiltonian (\ref{EHam})
\begin{equation}
\label{EHamsound}
\begin{aligned}
H[\varphi_q] =\frac{1}{2}\sum_{0\leq\omega_q^{l,t,t}\leq\omega_{\rm D}}[\pi_q^{1}\pi_{-q}^{1}+\pi_q^{2}\pi_{-q}^{2}+\pi_q^{3}\pi_{-q}^{3}]+\\
\sum_{0\leq\omega_q^l\leq\omega_{\rm D}}\left[\frac{\omega_q^2}{2}\varphi^1_q\varphi^1_{-q}\right]+ \sum_{\omega_{\rm F}\leq\omega_q^{t,t}\leq\omega_{\rm D}}\left[\frac{\omega_q^2}{2}(\varphi^2_q\varphi^2_{-q}+\varphi^3_q\varphi^3_{-q})\right]
\end{aligned}
\end{equation}
and
\begin{equation}
\omega_{\rm F}=\sqrt{\frac{\sigma^2}{\theta}}=\frac{2\pi}{\tau}=\frac{2\pi G_{\infty}}{\eta}
\label{ff}
\end{equation}
where the first term in Eq. (\ref{EHamsound}) is the contribution from $kinetic$ terms of both longitudinal and transverse phonon polarisations, the second is the $potential$ term from longitudinal phonon excitations and the third one is $potential$ contribution from two transverse phonon polarisations. At low frequencies ($0\leq\omega_q\leq\omega_{\rm F}$) transverse phonon modes are confined and, therefore, a sound can travel in the longitudinal direction only. Speed of sound in this low frequency range can be estimated in accordance with the Newton-Laplace equation
\begin{eqnarray}
\label{lc}
c_l=\sqrt{\frac{K}{\varrho}} \\
c_t=0
\end{eqnarray}
where $K$ is the bulk modulus and $\varrho$ is the density. At high frequency regime ($0\neq\frac{2\pi G_{\infty}}{\eta}=\omega_{\rm F}\leq\omega_q\leq\omega_{\rm D}$) the number of supported phonon modes both longitudinal and transverse is comparable to the number of modes in isotropically disordered solid. There is a non-zero stiffness both for volumetric deformations and shear deformations in liquids at high frequencies. Hence, it is possible to generate sound waves with different velocities dependent on the deformation mode. Sound waves generating volumetric deformations (compression) and shear deformations (shearing) are called pressure waves (longitudinal waves) and shear waves (transverse waves), respectively. The sound velocities of these two types of waves propagating in a homogeneous liquid, similarly to a disordered solid, can be estimated as
\begin{eqnarray}
\label{hc}
c_{l}= \sqrt{\frac{K+\frac{4}{3}G_{\infty}}{\varrho}} \\
c_{t}= \sqrt{\frac{G_{\infty}}{\varrho}}
\end{eqnarray}

Water has a second sound, also called "fast sound" \cite{bosse}, concerning the speed of sound. Over a range of high frequencies ($q>$ 4 $nm^{-1}$) liquid water behaves as though it is a disordered solid and sound travels at about twice its normal speed ($\sim$ 3200 $m s^{-1}$) \cite{ruocco1}, similar to the speed of sound in crystalline ice. The fast sound branch arises from the evolution of the "normal" sound mode ( see Eq. (\ref{lc})) at high frequencies \cite{ruocco2} where the restoring force of shear deformations acquires non-vanishing values (due to $G_{\infty}\neq 0$) and longitudinal velocity can be estimated through Eq. (\ref{hc}). Therefore, the transition of sound propagation in liquid water from $normal$ to {\it fast sound} regime is attributed to the transition from low frequency ($0\leq\omega_q\leq\omega_{\rm F}$) to high frequency  ($0\neq\frac{2\pi G_{\infty}}{\eta}=\omega_{\rm F}\leq\omega_q\leq\omega_{\rm D}$) regime
\begin{eqnarray}
\label{cl}
c_{l}= \sqrt{\frac{K}{\varrho}} \xrightarrow{q \rightarrow high \ q} c_{l}= \sqrt{\frac{K+\frac{4}{3}G_{\infty}}{\varrho}} \\
c_{t}=0 \xrightarrow{q \rightarrow high \ q} c_t=\sqrt{\frac{G_{\infty}}{\varrho}} 
\label{ct}
\end{eqnarray} 
which is well consistent with the phonon theory of liquids' picture \cite{bolsrph}. Moreover, this prediction was confirmed in inelastic X-ray experiment on deeply supercritical Ar and  a compelling evidence for the adiabatic-to-isothermal longitudinal sound propagation transition was provided \cite{boljpcl2015}.  Thus, a universal link is established between the positive sound dispersion (PSD) effect and the onset of shear phononic excitations revealing the viscous-to-elastic crossover in liquids and supercritical fluids. The PSD and shear sound propagation evolve consistently with theoretical predictions (see Eq. \ref{EHamsound} or \cite{sci1}). The simultaneous disappearance of both these effects at elevated temperatures is a manifestation of the Frenkel line, therefore, both can be considered as a universal fingerprint of the dynamic response of a liquid/fluid.

As we mentioned previously, liquids behave like isotropically disordered solids in terms of both longitudinal and transverse phonon excitations and have similar structure at elevated frequencies \cite{bolsrph,bolstr1,bolstr2}. Liquids sustain medium range order correlations but the translationally  invariant symmetry is broken \cite{sci1}. Therefore, liquids do not have periodicity and the fist and second Brillouin zones are not well defined as a result. The broken translational symmetry due to boundary smearing  between Brillouin zones and the transition we described above (see Eqs. (\ref{cl})-(\ref{ct})) leads to complete or partial phonon mode mixing at high frequency regime \cite{cunsolo,cunsolo1,cunsolo2,cunsolo3}. 

\subsubsection{Phonon localisation above the Frenkel line on P-T diagram}
Phonons are analogous to photons, having energy of $\hbar\omega$ as quanta of excitation of the lattice vibration mode of angular frequency $\omega$. Since the momentum $\hbar q$ is exact, by the Heisenberg's uncertainty principle, the position of phonons cannot be determined precisely. Hence, phonons are not localized particles in this sense. However, just like the case with photons or electrons, a fairly localized wave-packet can be constructed by combining modes of slightly different frequency and wavelength. By taking waves with a spread of $q$ of order $\frac{\pi}{na}$, a wave-packet localized within about $n$ unit cells is made, representing a fairly localized phonon with group velocity $\frac{\partial\omega_{q}}{\partial q}$, within the limits of the uncertainty principle. 

In translationally invariant systems phonons are always delocalised meaning that plane-waves easily travel at a given frequency range and localized states appear only near band edges implying that group velocity
\begin{equation}
v_{g}\equiv\frac{\partial\omega_{q}}{\partial q}\longrightarrow 0
\end{equation}
In {\it rigid liquid} regime the localisation of phonon modes $\left(\frac{\partial\omega_{q}}{\partial q}=0\right)$ occurs in transverse directions in the $0\leq\omega_{q}\leq\omega_{\rm F}$ frequency range, see last term in Eq. (\ref{EHamsound}). Localisation-delocalisation process in liquids happens at characteristic vibrational times $\tau\simeq\tau_{\rm F}$. By lowering the temperature in a rigid liquid we may observe delocalisation of phonon states at the lower transverse phonon band edge what is defined by the Frenkel frequency $\omega_{\rm F}=\frac{2\pi}{\tau}=\frac{2\pi G_{\infty}}{\eta}$, where $\eta=\eta_{0}\exp{\left(\frac{A}{T-T_{0}}\right)}$ according to VFT law \cite{bolprb}. Therefore, in liquids low-frequency transverse phonon modes are exponentially localised within a given temperature range. By increasing the temperature in a rigid liquid we may observe localisation of phonon states at the higher transverse phonon band edge. The transverse phonon propagation in a rigid liquid will be eliminated by reaching the Frenkel line thermodynamic limit 
\begin{equation}
\omega_{\rm F}\xrightarrow{T}\omega_{\rm D} \Longrightarrow c_V\xrightarrow{T}2k_{\rm B}
\end{equation}

\subsubsection{Phonon localisation below the Frenkel line on P-T diagram}
Once a liquid system reached the Frenkel line thermodynamic limit the transverse phonon propagating modes are not supported any more. This is valid both for monatomic and complex liquids. Nevertheless, in complex liquids such as supercritical carbon dioxide (CO$_2$) may occur phonon localisation effects on an intermediate length scale. Supercritical CO$_2$ below the Frenkel line has a peculiar shell structure where, in the first coordination shell, both carbon and oxygen atoms experience gas-like correlations with short-range
order interactions while within the second coordination shell, oxygen atoms essentially exhibit a liquid-like  correlations with medium-range order interactions due to localization of transverse-like phonon packets \cite{bolstr2}. Atoms inside of the nearest-neighbour heterogeneity shell play a catalytic role due to short-range
order correlations, providing a mechanism for diffusion in the supercritical CO$_2$ on an intermediate length scale \cite{bolstr2}. 

\begin{table*}
\caption{Structural and Dynamic properties of solids, liquids and gases, followed by two regimes in the supercritical phase: Rigid Liquid and Non-Rigid  Fluid. L and T letters refer to longitudinal and transverse phonon modes, respectively.}
\centering
\begin{tabular}{c c c}
\hline\hline
State &  Dynamics & Structure\\ [0.5ex]
\hline
{\begin{tabular}{ll}Solid\end{tabular}}  & 1-L+2-T Phonon Modes & Long and/or  Medium-Range Order \\
{\begin{tabular}{ll}Liquid\end{tabular}}   & 1-L+2-T ($\omega_{q} \geq\omega_{\rm F}$) Phonon Modes & Medium and/or  Short-Range Order\\
{\begin{tabular}{ll}Ideal \\ Gas\end{tabular}} & No Phonon Modes/No Interactions & No Order/Disorder \\
\hline\hline
{\begin{tabular}{ll}Rigid \\ Liquid\end{tabular}} & 1-L+2-T ($\omega_{q} \geq\omega_{\rm F}$) Phonon Modes & Medium-Range Order\\
\hline
{\begin{tabular}{ll}Non-Rigid \\  Fluid \end{tabular}}    & 1-L Phonon Mode & Short-Range Order \\
[1ex]
\hline\hline
\end{tabular}

\label{table1}
\end{table*}

The emergence of local-order heterogeneities (non-uniformities) and the existence of topologically confined transverse phonon modes (standing plane waves) on an intermediate length scale are closely interrelated. From equation (\ref{Ham}) we know that liquids support high-frequency propagating transverse phonon modes but lose this ability progressively as temperature increases extending well into the supercritical region. Oxygen atoms in the second coordination shell define a boundary between propagating and localized (topologically confined sound waves) phonon modes in the supercritical CO$_2$ \cite{bolstr2}, $c_V$: $2k_{\rm B}\xrightarrow{T}\frac{3}{2}k_{\rm B}$ \cite{bolnature}, where $c_V=\frac{3}{2}k_{\rm B}$ is the ideal gas thermodynamic limit. A 0.6 nm -- 0.8 nm  value is estimated as the length scale for highly localized transverse phonon modes which corresponds to the size of clusters where the phonon packets are localized. Similarly, the propagating lengths of the transverse acoustic modes were determined to be 0.4 nm--1 nm, corresponding to the size of cages formed instantaneously in liquid Ga \cite{hos1} and liquid Sn \cite{hos2}. Thus, the supercritical state is also amenable to supporting non-propagating (localised) transverse phonon excitations on an intermediate length scale.

\section{Equivalent approaches to liquid energy}
A rigorous mathematical description of liquid thermodynamics has always been regarded as much more difficult than that of the kinetic theory of gases or collective displacements of crystalline solids \cite{llandau}. It is now known that liquids do not have a simple interpolated status between a gas and a solid phases, nevertheless, similarities to the properties of both adjacent phases can certainly be observed \cite{dbolmatovjap}. The coefficients of self-diffusion of liquids (10$^{-5}$ cm$^2$s$^{-1}$) and solids (10$^{-9}$ cm$^2$s$^{-1}$) are orders of magnitude below those of gases. Viscosities of fluids and gases are some 13 orders of magnitude lower than those of solids, which we may easily understand in terms of momentum exchange. In terms of vibrational states, liquids differ from solids because they cannot support static shear stress ($G_{0}=0$). However, liquids support shear stress at high frequencies ($G_{\infty}\neq 0$). Flow in a solid arises primarily from rupturing of
bonds and the propagation of dislocations and imperfections. In a liquid, flow is characterized by both configurational and kinetic processes, whilst in a gas the flow is understood purely in terms of kinetic transport. In this very limited sense, liquids may have a minor partial interpolated status between gases and solids \cite{dbolmatovjap}.

\subsection{Dynamic approach based on contributions from phonon excitations}

\subsubsection{ Above the Frenkel line on the P-T diagram}
Here, we recall the analytical expression for the liquid energy derived earlier (see Eq. (\ref{E}))
\begin{equation}
E_{ph}=NT\left(1+\frac{\alpha T}{2}\right)\left(3D\left(\frac{\hslash\omega_{\rm D}}{k_{\rm B}T}\right)- \left(\frac{\omega_{\rm F}}{\omega_{\rm D}}\right)^{3}D\left(\frac{\hslash\omega_{\rm F}}{k_{\rm B}T}\right)\right)
\label{E1}
\end{equation}
\subsubsection{Below the Frenkel line on the P-T diagram}
Analogously, we represent the formula for liquid energy below the Frenkel line on the pressure-temperature diagram recalling the analytical expression (see Eq. (\ref{fen}))
\begin{equation}
E_{ph}=\frac{3}{2}Nk_{\rm B}T+\frac{3}{2}\left(1+\frac{\alpha T}{2}\right)Nk_{\rm B}T\frac{a^3}{\lambda^3}
\label{E2}
\end{equation}
The comparison of both dynamic (Eqs. (\ref{E1})-(\ref{E2})) and structural (Eq. (\ref{E3})) approaches to liquid energy we discuss below.

\subsection{Structural approach based on contributions from pair correlations}
Alternatively, internal energy $E$ of a system can be calculated by computing the pair distribution function 
$g(r)$. Here, we consider the spherical shell of volume $4\pi r^{2}\delta r$ which contains $4\pi \varrho g(r)\delta r$ particles, where $\varrho=\frac{N}{V}$ is the number density. If the pair potential at a distance $r$ has a value $v(r)$, the internal energy  of interaction between the particles in the shell and the central particle is $4\pi r^{2}\varrho g(r)v(r)\delta r$. The total energy of $N$ interacting particle is given as
\begin{equation}
E_{str}=\frac{3}{2}Nk_{\rm B}T+2\pi N\varrho\int_{0}^{\infty}r^{2}v(r)g(r)dr
\label{E3}
\end{equation}
and the pair distribution function $g(r)$ can be expressed through the static structure factor $S(q)$
\begin{equation}
\label{gor}
g(r)=1+\frac{1}{2\pi\varrho}\int_{0}^{\infty}[S(q)-1]\frac{\sin{qr}}{qr}q^{2}dq
\end{equation}
which can be measured in diffraction experiments.

In general, both dynamic (based on phonon contributions, see Eqs. (\ref{E1})-(\ref{E2})) and structural (based on contributions from pair correlations, see Eq. (\ref{E3})) approaches to internal energy and heat capacity can be considered as equivalent. However, these approaches operate differently considering specific thermodynamic limits. Regarding solids, approaching the quantum limit at low temperatures the dynamic approach easily accounts for quantized portions of energy and predicts low-$T$ thermodynamic regime for solids ($c_{V}\sim T^{3}$)( for detail see subsection {\it Low-T quantum limit}). In contrast, the structural approach has a limitation describing solids at low temperatures. Even, in case of weak interactions the second term in Eq. (\ref{E3}) occurs negligible and the minimum achievable heat capacity becomes $c_V=\frac{1}{N}\left(\frac{\partial E_{str}}{\partial T}\right)_{\rm V}\xrightarrow{T}\frac{3}{2}k_{\rm B}$ reaching the ideal gas thermodynamic limit only, far beyond the quantum limit though. Nevertheless, both $dynamic$ and $structural$ approaches are interconnected over a wide range of pressure and temperature on the phase diagram (see Table III) and it can also be observed through calculation the Green-Kubo transport coefficients.

\section{Green-Kubo relations}
Green and Kubo showed \cite{green,kubo} that phenomenological coefficients like  self-diffusion coefficient, thermal conductivity  and shear viscosity could be written as integrals over a time-correlation function in thermal equilibrium. Lars Onsager stated in his regression hypothesis \cite{lars} that fluctuations are present in every equilibrated system. Dissipation of fluctuations has the same origin as the relaxation toward equilibrium once the system is out of equilibrium due to an external force. Both the dissipation and the relaxation time scales are determined by the same transport coefficients.

In the standard linear response theory for a system subject to an external field, the Green-Kubo formula for a linear response coefficient can be expressed in terms of time- and/or distance dependant correlation function. Let us consider time-dependant 
\begin{equation}
\label{ct}
C(t)=\langle C(t_{0})C(t_{0}+t)\rangle
\end{equation}
and distant-dependant 
\begin{equation}
\label{cd}
C(\vec{r})=\langle C(\vec{r}_{0})C(\vec{r}_{0}+\vec{r})\rangle
\end{equation}
 correlation functions, where $\langle\rangle$ represents the average over initial conditions. Function Eq. (\ref{ct}) is not a function of time but the function of the shift in time or correlation time. The same is applicable also to correlations in space (see Eq. (\ref{cd})), instead of time. 
 
Vibrational spectrum for a system can be calculated by taking the Fourier transform of the correlation function, $C(\tau)$ that transfers the information on the correlations along the atomic trajectories from time to frequency frame of reference
\begin{equation}
I(\nu)=\int_{-\infty}^{\infty}\exp{(-2\pi i\nu\tau)}C(\tau)d\tau
\end{equation}
where $C(\tau)$ is the velocity-velocity autocorrelation function for an ensemble of N particles 
\begin{equation}
C(\tau)=\frac{1}{N}\sum_{i=1}^{N}\frac{1}{t_{max}}\sum_{t_{0}=1}^{t_{max}}\vec{v}_{i}(t_{0})\cdot\vec{v}_{i}(t_{0}+\tau)
\end{equation}
Therefore, the vibrational spectrum of a system 
\begin{equation}
\label{I1}
I(\omega)\propto\langle\vec{v}(0)\vec{v}(t)\rangle
\end{equation}
On the other hand, the structural spectrum can be measured in neutron and X-ray scattering experiments
\begin{eqnarray}
I(q)=S(q)\sum_{i=1}^{N}f_{i}^{2}(q) \Longrightarrow \\
I(q)\propto [g(r)-1]
\label{I2}
\end{eqnarray}
because of Eq. (\ref{gor}) and $f(q)$ is the form factor. In general, combining results from Eqs. (\ref{I1}) and (\ref{I2}) one may write
\begin{equation}
I(q,\omega)\propto I(q)\otimes I(\omega)
\end{equation}
mathematically repeating the reconstruction of experimental measurement procedure. Hence, both vibrational and structural correlations can be studied independently in temporal and spatial domains, respectively. As dynamic and structural approaches are  interlinked and complementary, change in structure with temperature variations leads to change in dynamics and $vice$ $versa$, which is consequence of the equation below 
\begin{equation}
E_{phonons}=E_{pair \ correlations}
\end{equation}
therefore, dynamic (phonon approach) and structural (pair correlations approach) properties in condensed phases change consistently with temperature variations. Below we test this general statement studying evolution of correlation functions within the Green-Kubo phenomenological transport coefficients formalism. To calculate these correlation functions, we have used LAMMPS simulation code to run a Lennard-Jones (LJ, $\epsilon/k_{\rm B}$=119.8 K, $\sigma$=3.405) fluid fitted to Ar properties \cite{ljpot} with 32678 atoms in the isothermal-isobaric
(NPT) ensemble. 
\subsection{Velocity-velocity autocorrelation function: temporal correlations}
An explicit calculation of the velocity-velocity autocorrelation function (VACF) and of its frequency spectrum for liquid Argon in particular has been well studied \cite{rah1,rah2,alder1,alder2}. If the velocity vector for a system of atoms is $\mathbf{v}(t)$, then the velocity autocorrelation can be written as
\begin{equation}
\displaystyle C(t) =\frac{\left\langle\mathbf{v}(0)\cdot\mathbf{v}(t)\right\rangle}{\left\langle\mathbf{v}(0)\cdot\mathbf{v}(0)\right\rangle}
\end{equation}
and $D\propto C(t)$, where $D$ is the diffusion constant. In a liquid the VACF is not a simple exponential but has a negative part as its characteristic feature. The decay of the correlation function, in a liquid in particular, reflects the decay in the correlations in atomic motion along the trajectories of the atoms, not in the amplitude. $C(t)$ has the long time tail which can be explained as a collective dynamic effect: part of the momentum of a particle is stored in a microscopic vortex that dies off very slowly \cite{dorf}.
\begin{figure}
	\centering
\includegraphics[scale=0.35]{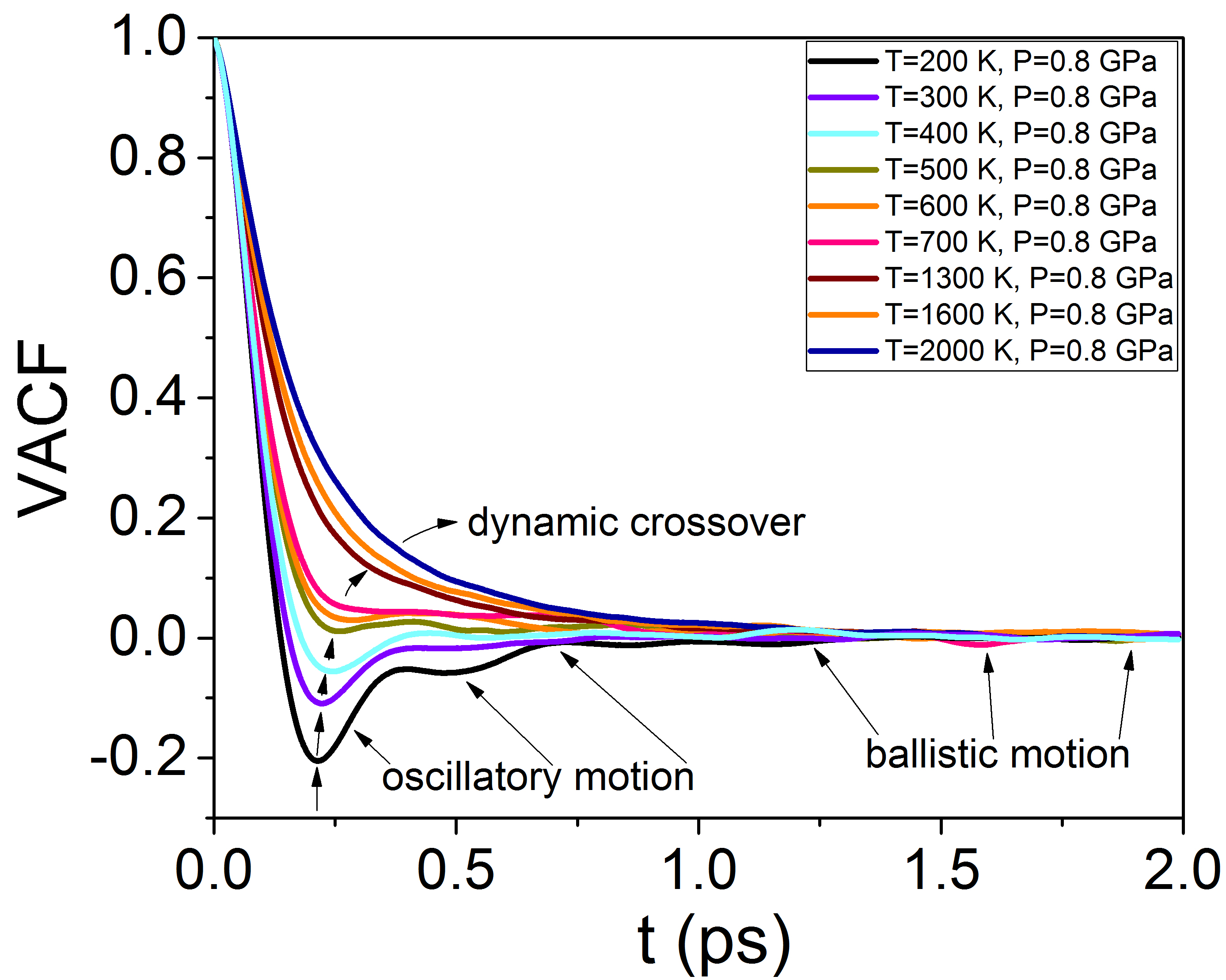}
\caption{Temperature variations of the velocity-velocity autocorrelation function at fixed pressure P=0.8 GPa. VACF exhibits oscillatory motions up to 1-1.5 ps and ballistic 
motions beyond 1.5 ps time scale. VACF undergoes the dynamic crossover upon crossing the dynamic boundary between T=400 K (oscillatory motions almost disappeared) and T=500 K (motions are almost purely ballistic).}
	\label{vacf}
\end{figure}
If the motion of the atoms tends to an oscillatory pattern, the VACF will characterize the ossilations because the velocity of the atoms will self-correlate in a periodic fashion. Similarly, if the velocity of atoms tends to one direction, then the VACF will gradually grow in magnitude characterizing the diffusion of the atoms. In Fig. (\ref{vacf}) we plot the evolution of VACF with temperature variations at fixed pressure for liquid Ar and evidence the dynamic crossover: the continuous transition from oscillatory to ballistic-collisional motion respectively \cite{brazhkinprl}. 
\subsection{Pair distribution function: spatial correlations}
\begin{figure}
	\centering
\includegraphics[scale=0.35]{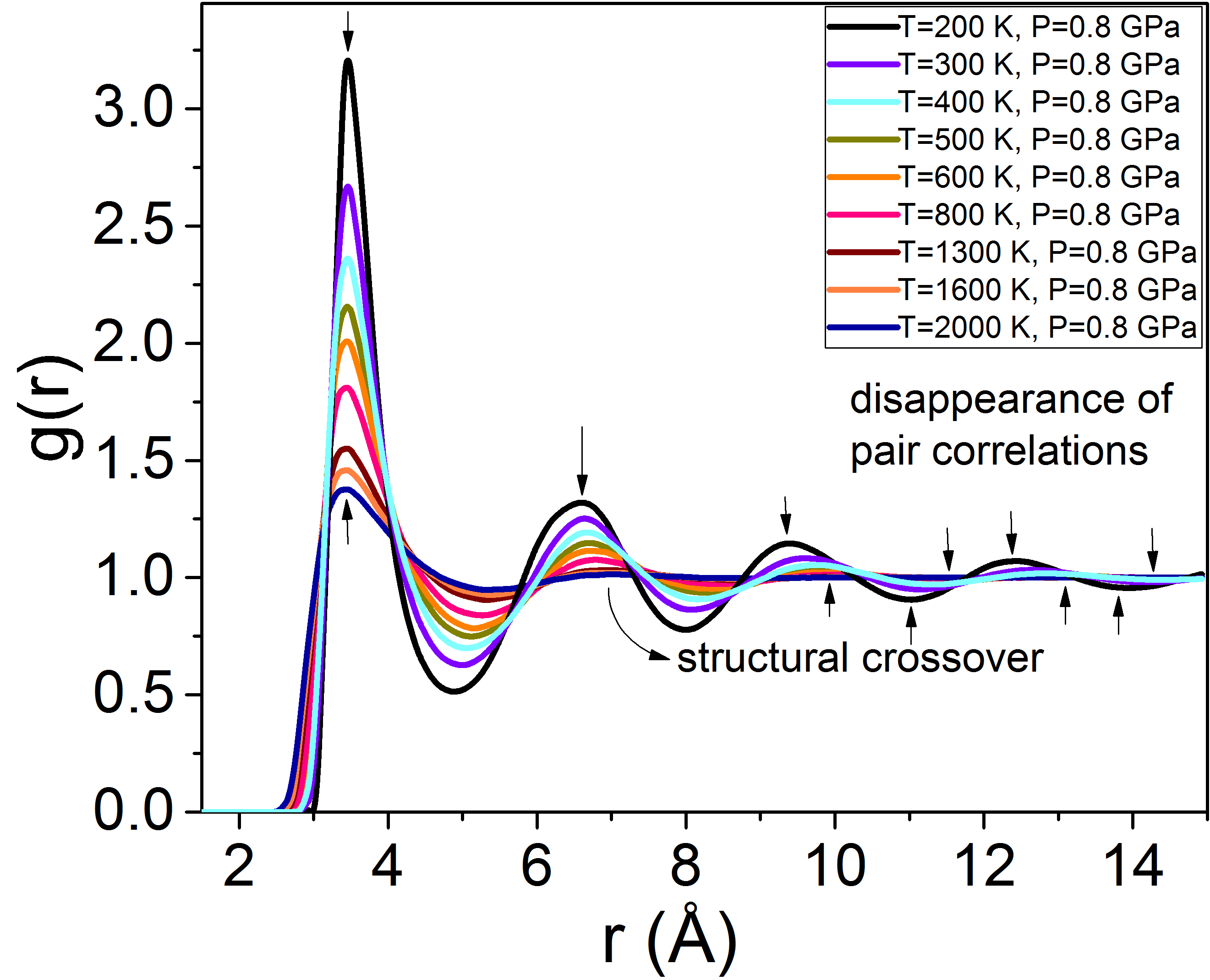}
\caption{Temperature variations of the pair correlation function $g(r)$ at fixed pressure P=0.8 GPa. The $g(r)$ functions of simulated one-component Lennard-Jones (LJ) fluid at different temperatures showing the
disappearance of the medium-range order upon crossing the structural boundary in the supercritical state between T=400 K and T=500 K.}
	\label{pdf}
\end{figure}

Distant-dependant correlation function $C(\vec{r})$, see Eq. (\ref{cd}), can be expressed as (also see \cite{bolsph})
\begin{equation}
C(\vec{r})=\frac{1}{N\varrho}\sum_{ij}^{N}\delta(\vec{r}-\vec{r}_{ij})
\end{equation}
For isotropic system $g(r)$ can be averaged over angles by calculating an average number of particles at distances $r$ and $r+\Delta r$ 
\begin{equation} 
g(r)=\frac{\langle C(\vec{r})\rangle}{4\pi r^{2}\Delta r}
\end{equation}
In Fig. (\ref{pdf}) we show the evolution of $g(r)$ with temperature variations at fixed pressure for liquid Ar and evidence of the structural crossover: the continuous transition from medium-range order pair correlation to short-range order pair correlation respectively \cite{bolstr1,bolstr2}. 
The structural crossover in the supercritical state is the new effect (see Fig. \ref{pdf}), in view of the currently perceived physical homogeneity of the supercritical state. Here, we show that a liquid Ar undergoes a structural crossover which corresponds to the qualitative change of atomic structure, the transition of the substance from the "rigid" liquid structure (pair correlations presented at an intermediate length scale) to the "non-rigid" fluid structure ((pair correlations preserved at a short length scale only).

It is noteworthy, that any approach what bears a "small parameter" and makes it possible to expand the internal energy of the system into the series such as virial expansion, cluster expansion, Meyer expansion, Percus-Yevick approach and hard-sphere model is only capable of explaining weakly interacting systems \cite{jpercus}. However, all these approaches fail in describing the systems with strong interactions like $rigid$/compressed liquids. Hence, these approaches are unable to predict and describe the structural crossover in the supercritical state \cite{bolstr1,bolstr2,bolf}.

We now address the origin of the structural crossover (see Fig. \ref{pdf}), and relate this origin to the changes of both dynamics (see Fig. \ref{vacf}) and thermodynamics (see Tables \ref{table}-\ref{table1}) of the liquid matter. Indeed, $g(r)$ peaks (their heights and positions) on both sides of the structural crossover  change differently with temperature increase.  In the {\it rigid liquid} regime, $g(r)$ peaks decrease rapidly due to the exponential decrease of the relaxation time (also see \cite{bolstr1}). In the {\it non-rigid fluid} regime where the oscillatory component of motion (see Fig. \ref{vacf}) and the medium-range order pair correlations are no longer present (see Fig. \ref{pdf}), pair correlations are less sensitive to temperature increase because the dynamics is already randomized by ballistic motions as in a gas meaning that the system reached the Frenkel line thermodynamic limit $c_V=2k_{\rm B}$. This picture supports the relationship between the
structure, dynamics and thermodynamics and its consistent temperature evolution over a wide range of temperature and pressure on the phase diagram (also see Tables \ref{table}-\ref{table1}). 

\section{Discussions and Conclusions}
The interatomic interactions in a liquid are strong, and therefore firmly affect calculations of its thermodynamic properties. At the same time,  liquids, occupy an interesting intermediate state with a combination of strong interactions and cohesive state as in solids and large flow-enabling particle displacements as in gases, the calculation of liquid energy in particular requires the explicit knowledge of the interactions. For this reason, it is argued that no general expressions for liquid energy can be obtained, in contrast to gases and solids \cite{llandau}. An apt summary of this state of affairs, attributed to Landau, is that liquids "have no small parameter." Perhaps for this reason, liquid energy and heat capacity have not, or barely, been mentioned in statistical physics textbooks as well as books dedicated to liquids \cite{ziman,boon,march,hansen}. This is believed have precluded the calculation of thermodynamic properties of liquids in general form. Nevertheless, in the last few decades a great amount of activity has taken place in the field of statistical mechanics of disordered materials and liquids in particular. This activity was mainly originated by the development of various theories and novel experiments with which the predictions of these theories could be tested. One of the most outstanding examples of these theories are the mode-coupling theory (MCT) \cite{gotze,leu} and the fundamental-measure theory by Rosenfeld and Tarazona which gives a unified analytical description of classical bulk solids and fluids \cite{rosenfeld}, and describe scaling in fluids \cite{dyre13}.
 
In this work, we propose an unified approach to states of aggregation existing on the P-T phase diagram: solid, liquid and gas phases. We derive the effective Hamiltonian with low-energy cutoff in two transverse phononic polarisations as a result of symmetry breaking in phonon interactions. This result rigorously proves the Frenkel's intuitive microscopic picture about phonon excitations in condensed bodies \cite{ffrenkel} on the fundamental basis. Further, we construct the statistical mechanics of states of aggregation existing on the P-T phase diagram by employing the Debye approximation. The introduced formalism also covers the well-known thermodynamic limits such as the Delong-Petit thermodynamic limit ($c_V=3k_{\rm B}$) and the ideal gas limit ($c_V=\frac{3}{2}k_{\rm B}$). In the framework of the proposed formalism we derive the new thermodynamic limit $c_V=2k_{\rm B}$, dubbed here the Frenkel line thermodynamic limit. In the framework of the phonon theory of liquids \cite{bolsrph}, which we derive from the effective Hamiltonian (\ref{EHam}), we discuss the phonon propagation and localisation effects in liquids above and below the Frenkel line, and explain the "fast sound" phenomenon. As a test for our theory we calculate velocity-velocity autocorrelation and pair distribution functions using the linear response phenomenological formalism (introduced by Green and Kubo). As a result, we show the consistency between dynamics, structure and thermodynamics on the P-T phase diagram in the framework of the introduced unified approach. In particular, we prove that dynamic, structure and thermodynamics evolve consistently upon crossing the Frenkel line. The existence of the Frenkel line (can be predicted from Eq. (\ref{EHam})) is important for studies fluids under extreme conditions of pressure and temperature. Recently, it was shown that in Jupiter and Saturn, supercritical molecular hydrogen undergoes a smooth dynamic transition around 10 GPa and 3000 K from the {\it rigid liquid}  regime to the {\it non-rigid fluid} regime upon crossing the  Frenkel line \cite{trachpre}. The continuous transition is accompanied by qualitative changes of all major physical properties. The implication of that work provides a physically justified way to demarcate the interior and the atmosphere in gas giants \cite{trachpre}. A similar work was done on supercritical carbon dioxide providing a compelling evidence for a structural crossover \cite{bolstr1} in the supercritical state \cite{bolstr2}. An intriguing question whether Venus may have had CO${_2}$ oceans was raised urging experimental groups for the verification of the proposed hypothesis \cite{bolstr2}.

From Eq. (\ref{EHam}) we derive the phonon theory of liquid thermodynamics \cite{bolsrph} (in the {\it rigid liquid} regime section) which has successfully been applied to calculating various thermodynamic properties in nonofluidics \cite{saito} and at solid/liquid interfaces \cite{hopkins1,hopkins2} by other research groups. Phononic band gaps manipulation (see the last term in Eq. (\ref{EHam})) is perhaps the most obvious application in a scope of sound control technologies. Sound is immensely valuable in our daily lives for communication and information transfer. The existence of liquid structures with complete phononic band gaps has obvious applications. For example, a
phononic crystals immersed in a liquid or aqueous solution will reflect incoming sound waves with frequencies within the gap and can therefore be used as an acoustic insulator. Moreover, the introduction of
defects within hybrid structures (dissolved nano-clusters in liquids) allows sound waves with frequencies in the band gap to be trapped near a point-like defects and/or guided along linear
defects. Therefore, we expect that the reported results in this work will advance technologies based on the THz phononic band gaps control.

\section{Acknowledgements} 
This work was partly supported by the U.S. Department of Energy, Office of Science, Office of Basic Energy Sciences, under Contract No. DE-SC0012704. We are indebted to Stefano Ruffo, Yang Zhang, Salvatore Torquato, Oleg Kogan, Ivar Martin, Yugang Zhang and Oleg Gang for stimulating discussions.

\end{document}